\documentclass{aa}
\usepackage{graphicx}

\usepackage{url}
\usepackage{kantlipsum} 

\usepackage[colorlinks=true]{hyperref}
\hypersetup{citecolor= blue, linkcolor=red}
\usepackage{txfonts}

\usepackage{float}

\usepackage{natbib}
\bibpunct{(}{)}{;}{a}{}{,} 

\newcommand{\NICOLE}{{\sc NICOLE}}
\newcommand{\Sir}{{\sc Sir}}

\newcommand{\kms}{\,km\,s$^{-1}$}

\newcommand{\CHROTEL}{{\sc Chrotel}}
\newcommand{\BIFROST}{{\sc Bifrost}}
\newcommand{\dyn}{\,dyn\,cm$^{-2}$}
\usepackage{cancel}

\begin{document} 

   \title{Diagnostic potential of the \ion{Ca}{ii}~8542\,\AA\ line for solar filaments}


   \author{C. J. D\'iaz Baso
          \inst{1,2,3}
          \and
          M. J. Mart\'{\i}nez Gonz\'alez
          \inst{1,2}
          \and
          A. Asensio Ramos
          \inst{1,2}
          \and
          J. de la Cruz Rodr\'iguez
          \inst{3}
          }

   \institute{Instituto de Astrof\'isica de Canarias, C/V\'{\i}a L\'actea s/n, E-38205 La Laguna, Tenerife, Spain
   \and
   Departamento de Astrof\'{\i}­sica, Universidad de La Laguna, E-38206 La Laguna, Tenerife, Spain
   \and
   Institute for Solar Physics, Dept. of Astronomy, Stockholm University, AlbaNova University Centre, SE-10691 Stockholm Sweden
             }

   \date{Received Month Day, 2018; accepted Month Day, 2018}
   
   \authorrunning{D\'iaz Baso et al.}
 
  \abstract
   {}
   {In this study we explore the diagnostic potential of the chromospheric \ion{Ca}{ii} line at 8542\,\AA\ for studying the magnetic and dynamic properties of solar filaments. We have acquired high spatial resolution spectropolarimetric observations in the \ion{Ca}{ii}~8542 \AA\ line using the CRISP instrument at the Swedish 1-m Solar Telescope.}
   {We use the \NICOLE\ inversion code to infer physical properties {from observations} of a solar filament. We discuss the validity of the results due to the assumption of hydrostatic equilibrium. We have used observations from other telescopes such as \CHROTEL\ and SDO, in order to study large scale dynamics and the long term evolution of the filament.}
   {We show that the \ion{Ca}{ii}~8542 \AA\ line encodes information of the temperature, line-of-sight velocity and magnetic field vector from the region where the filament is located. The current noise level only allow us to estimate an upper limit of 260\,G for the total magnetic field of the filament. Our study also reveals that if we only consider information from the aforementioned spectral line, the geometric height, the temperature and the density can be degenerated parameters outside the hydrostatic equilibrium approach.}
   {}

   \keywords{Sun: filaments, prominences -- Sun: chromosphere -- Sun: magnetic fields -- Sun: oscillations --  Sun: evolution}

   \maketitle


\section{Introduction}\label{sec:intro}

\begin{figure*}[!ht]
\centering
\includegraphics[width=0.98\linewidth]{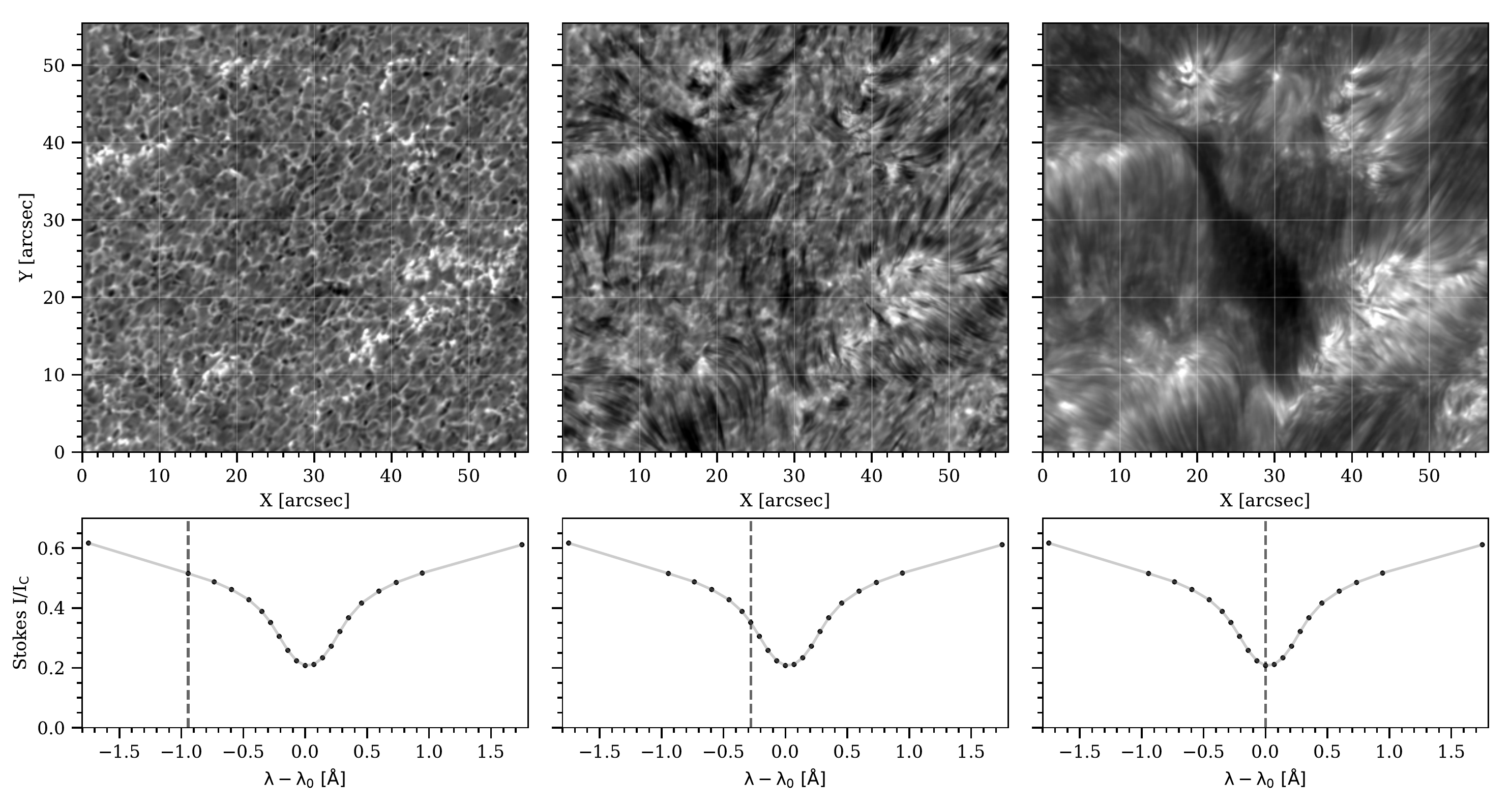}
\caption{FOV at three different wavelength positions $-0.945$, $-0.28$, and $0.0$\,\AA\ from the line center $\lambda_0=8542.1$\,\AA. The lower panels show an average line profile and a vertical dashed line at the wavelength used to generate the field of view of the upper panel. We see how the spectral line has information from the photosphere (left column) to the chromosphere (right column) where the filament is observed.}
\label{fig:observation}
\end{figure*}

Solar filaments are cool plasma overdensities suspended at chromospheric and coronal heights. Instead of falling to the surface due to gravitational acceleration, they can stay suspended from hours to several weeks. Forces derived from the presence of magnetic fields are expected to provide the required support to form filaments. However, the exact magnetic topology of filaments and its connectivity with the photosphere are still an on-going debate. Filament plasma is supported where the magnetic field lines are curved upwards, forming dips. {The latter} could be induced by the prominence plasma itself \citep{Kippenhahn1957}, or can be present in untwisted \citep{Antiochos1994}, twisted \citep{van1989} or tangled \citep{vanB2010} magnetic fields.

This controversy about the {magnetic field} topology is also connected with the dynamics of the filament. For example, streaming and counter-streaming flows have been identified in the spine and barbs of filaments interpreted as plasma flows along not dipped magnetic flux tubes  \citep{Zirker1998,Lin2003,Diercke2018}. However, other authors, such as \cite{heinzel2007}, {have shown} that analyzing 2D images can lead to misinterpretations because of the lack of a fully 3D dynamical pattern and the counter-streaming could be explained with oscillations rather than flows along threads.

Although many studies and simulations have been carried out in the past decades \citep{Aulanier2003,Lites2005,Khomenko2014RTI,Keppens2014,Luna2015,Okamoto2016}, new high-resolution observations are opening the door to a better understanding of this solar phenomenon \citep{marian2015,kuckein2016}. These chromospheric structures can be observed in the core of strong lines with sufficient opacity to be sensitive to the physical conditions above the photosphere. Most studies have used {observations acquired in} H$\alpha$ \citep{kuckein2016}, the \ion{He}{i} multiplets at 10830\,\AA\ \citep{kuckein2012,orozco2014} and 5876\,\AA\ \citep[D$_3$,][]{Leroy1977,LopezAriste2002,Casini2003}, and the extreme ultraviolet continua \citep{Schmieder2003} to study {filaments}.

Solar filaments can also be {observed} in the \ion{Ca}{ii} infrared triplet at 8498, 8542 and 8662\,\AA. The latter have relatively low effective Land\'e factors ($\bar{g}_{8498}$=1.07, $\bar{g}_{8542}$=1.10, and $\bar{g}_{8662}$=0.87). Although they are not especially sensitive {to the presence of magnetic fields}, these lines can be observed in the entire solar disk (as opposite to the \ion{He}{i}~10830\,\AA\ triplet, whose opacity in the quiet Sun is extremely {low}) and we can use them to extract information about the plasma temperature \citep{socas2000}. Non-LTE radiative transfer is required to model the \ion{Ca}{ii}~IR lines, in this case the assumption of statistical equilibrium and complete frequency and angle redistribution suffice to retrieve accurate results \citep{Uitenbroek1989,Wedemeyer2011}.

The \ion{Ca}{ii}~8542\,\AA\ has been most used in previous studies because it has the largest opacity among the triplet lines. Additionally, it has the largest Land\'e factor of the triplet and it is not blended like the \ion{Ca}{ii}~8498\,\AA. {The \ion{Ca}{ii} IR lines are very sensitive to the temperature stratification of the lower chromosphere and 3D radiative transfer effects can be neglected} \citep{delaCruz2012,Leenaarts2009M,Stephan2016}. Despite these advantages,
its low polarimetric sensitivity has limited the study of magnetic fields to active regions with large magnetic field {strengths} \citep{socas2000,2005ApJ...633L..57S,delaCruz2013,2015ApJ...810..145D,asensio2017}. Outside active regions the magnetic field is so weak that {scattering polarization and} the Hanle effect must be included{, especially in observations acquired towards the limb}. The Hanle regime of the \ion{Ca}{ii}~8542\,\AA\ line is between 0.001<$B$<0.1~G and the Hanle saturation regime for $B$>10~G \citep{manso2010}. Above 100~G the line is again sensitive to the magnetic field vector through the Zeeman effect. However, the presence of 3D and dynamic effects in the radiative transfer problem make it very time-consuming in terms of computation \citep{carlin2013,Stephan2016}.

There are only a few studies of prominences (off-limb {filaments}) in the \ion{Ca}{ii}~8542\,\AA\ line \citep{Stellmacher2003,Park2013} where they  calculate the temperature from the line broadening using two lines of different atoms to separate the contribution of the thermal and non-thermal motions, using \ion{He}{i}~10830\,\AA, H$\alpha$, and H$\beta$. For filaments (on-disk) the background, which is illuminating the structure,  needs to be taken into account and cloud models are used \citep[see the review by][]{Tziotziou2007}.

In this study we present the analysis of very high spatial resolution spectropolarimetric observations acquired in the  \ion{Ca}{ii} 8542\,\AA\ line which are analyzed using the NLTE inversion code \NICOLE\ \citep{socas2000,SocasNavarro2015}. In our analysis we explore the potential diagnostic of the \ion{Ca}{ii} line and we discuss the validity of inversion results due to the assumption of hydrostatic equilibrium. To study the global dynamics of the region and the evolution of the filament, we have used observations from other synoptic telescopes.

\section{Observations}\label{sec:observation}

\begin{figure*}[!ht]
\centering
\includegraphics[width=\linewidth]{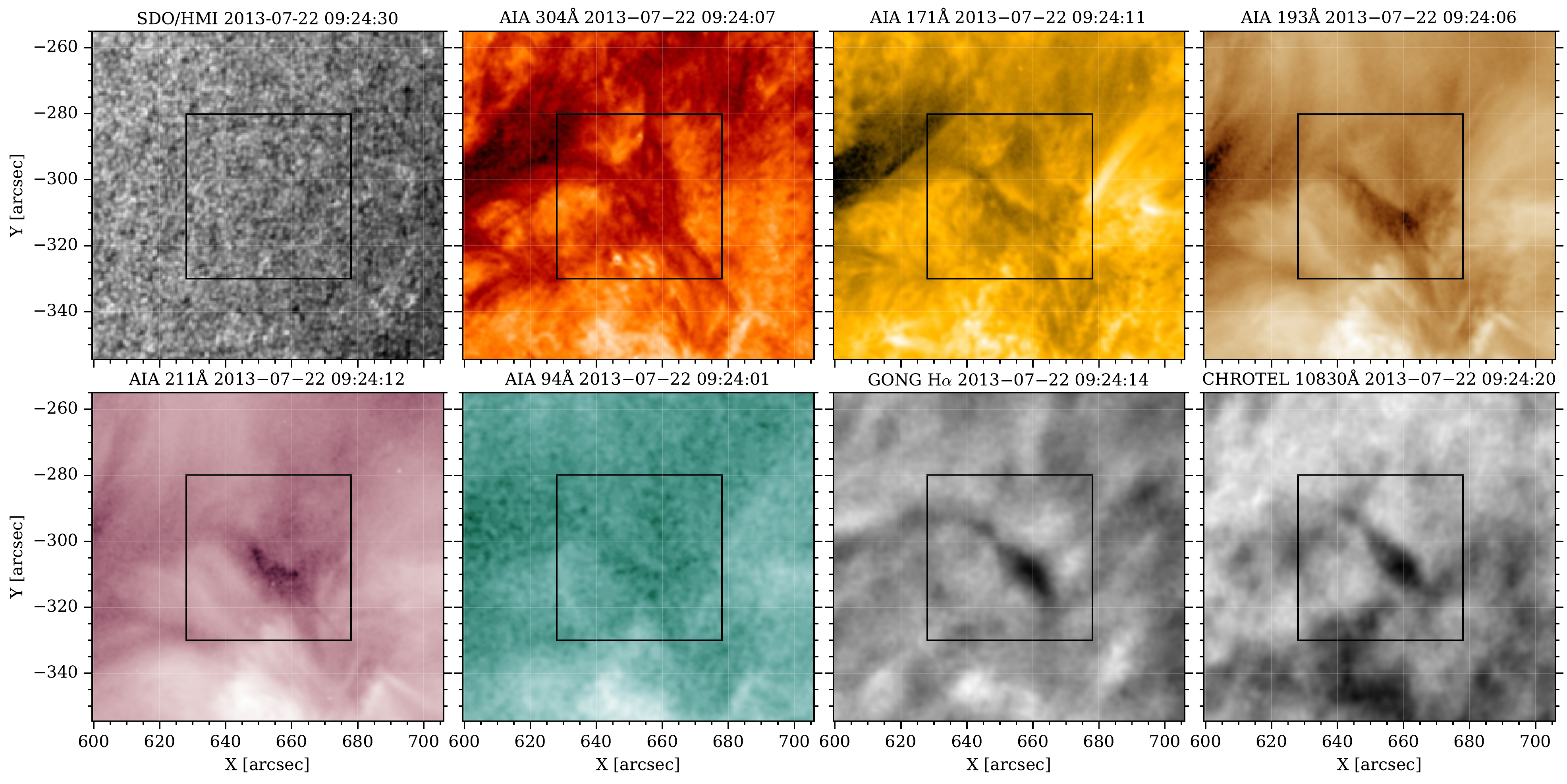}
\caption{We show an area of 100\arcsec$\times$100\arcsec, with a square displaying our FOV. The wavelength range at which each map is taken is indicated at the top of each panel. The axes scales have their origin at the disc centre, being the vertical the N-S direction while the horizontal the W-E direction, with negative sign towards the south and the west.}
\label{fig:sdomaps}
\end{figure*}


The observations analyzed in this study were {acquired} at the Swedish 1-m Solar Telescope \citep[SST,][]{scharmer2003} on the 22$^\mathrm{nd}$ of July 2013 using the CRISP {imaging spectropolarimeter} \citep{scharmer2008}. The telescope was pointed to a solar filament situated far from disk center at coordinates x, y = (+653, --305)\arcsec, which corresponds to {heliocentric distance} $\mu={\rm cos}\,\theta=0.65$.


The \ion{Ca}{ii} 8542\,\AA\ line was sampled in the range $\pm$1.75\,\AA\ from the core in 21 {line positions sampled with varying step size}: 70\,m\AA\ close to the core of the line, and up to 800\,m\AA\ in the far wings (see lower panels of Fig.~\ref{fig:observation}). This strategy \citep{delaCruz2012,delaCruz2013} is used to {optimize the time cadence of the observations by placing less spectral points in the far photospheric wings}. The observations consist of 36 scans taken with a cadence of $\sim$23.22\,s. {Simultaneous observations in the \ion{Fe}{i}~$6301/6302$\,\AA\ lines were also acquired but not used in the present study.}

Figure~\ref{fig:observation} shows the field of view (FOV) at different wavelength positions. We have chosen three different $\lambda-\lambda_0$ offsets at $-0.945$, $-0.28$, and $0.0$\,\AA\ from the line center in order to visualize how we are mapping different {atmospheric regimes as} we move across the spectral line. In the left panel, {the image depicts} reversed granulation at a height around 140\,km \citep{cheung2007} where the intergranular lanes are relatively hotter than the granules, in contrast to what is observed at the lower photosphere. This image gives us an idea of which are the deepest layers we can measure given that we do not reach continuum wavelengths. The middle panel shows a filamentary structure that covers the whole FOV. This structure is even more enhanced in the right panel, where bright plage areas are surrounding the dark filament showed in absorption in the core of the \ion{Ca}{ii} 8542\,\AA\ line. This filament will be the focus of this study.

\subsection{Data reduction process}
\label{sec:reductionProcess}

The reduction of the CRISP dataset was carried out using the CRISPRED pipeline \citep{delaCruz2015}. This software takes into account the dark current subtraction, flat-field correction, and demodulation process. The data have been processed with the MOMFBD code \citep{vanNoort2005} to remove blurring effects of atmospheric distortion while bringing the spatial resolution close to the diffraction limit of the telescope at this wavelength ($\sim$0.18''). The resulting noise level is in the range $4-6\cdot10^{-3}$ in units of the continuum intensity for Stokes $Q$, $U$, and $V$.

After the standard data reduction, a final inspection showed some residual artifacts that required to be corrected. The first one is the presence of polarized interference fringes. These fringes {appear} in the individual monochromatic images. The spatial pattern of these fringes is different for each Stokes parameter and it is also wavelength dependent. To correct these fringes we had to compute Fourier filters for each individual wavelength and Stokes parameter. The frequency amplitude of these fringes is well localized, far away from the frequencies involved in the data and a threshold is used to set those peak coefficients to zero (i.e., we do not use a rectangular or fixed shape to mask the fringe frequencies). Then, we do the inverse Fourier transform to recover our clean image. We note that these fringes are almost at the level of our signals {and a very careful estimation of the fringe pattern is required to properly retrieve the magnetic field vector}. After the cleaning process, though not completely removed, the presence of fringes was strongly alleviated.

For the correction of very-long period fringes or cross-talk, we have used monochromatic images at continuum wavelengths of each Stokes parameter. The advantage of continuum wavelengths (or very far wings in this case) is that we do not expect any significant structure in Stokes $Q$, $U$, or $V$. Therefore, the identification of this effect is easier and more accurate. {In order to apply this correction}, we assumed that every monochromatic image for each Stokes parameter is composed of the \emph{true} measurement contaminated by a quantity that is proportional to the signal detected in the line wing: 
\begin{equation}
X(\lambda)=X_0(\lambda) + C_{\mathrm{X,\lambda}} X_{\mathrm{W}}. 
\end{equation}
The coefficient $C_{\mathrm{X,\lambda}}$ is estimated by minimizing the absolute correlation between the far-wing image and each monochromatic image, and this contribution is then removed from the input image. Since measurements at different wavelengths on the continuum show slightly different patterns, we cannot ensure the accuracy of the polarization signals below or at the level of $6\cdot10^{-4}I_c$  as the artifacts have an amplitude of the same order.

\begin{figure*}[!ht]
\centering
\includegraphics[width=\linewidth]{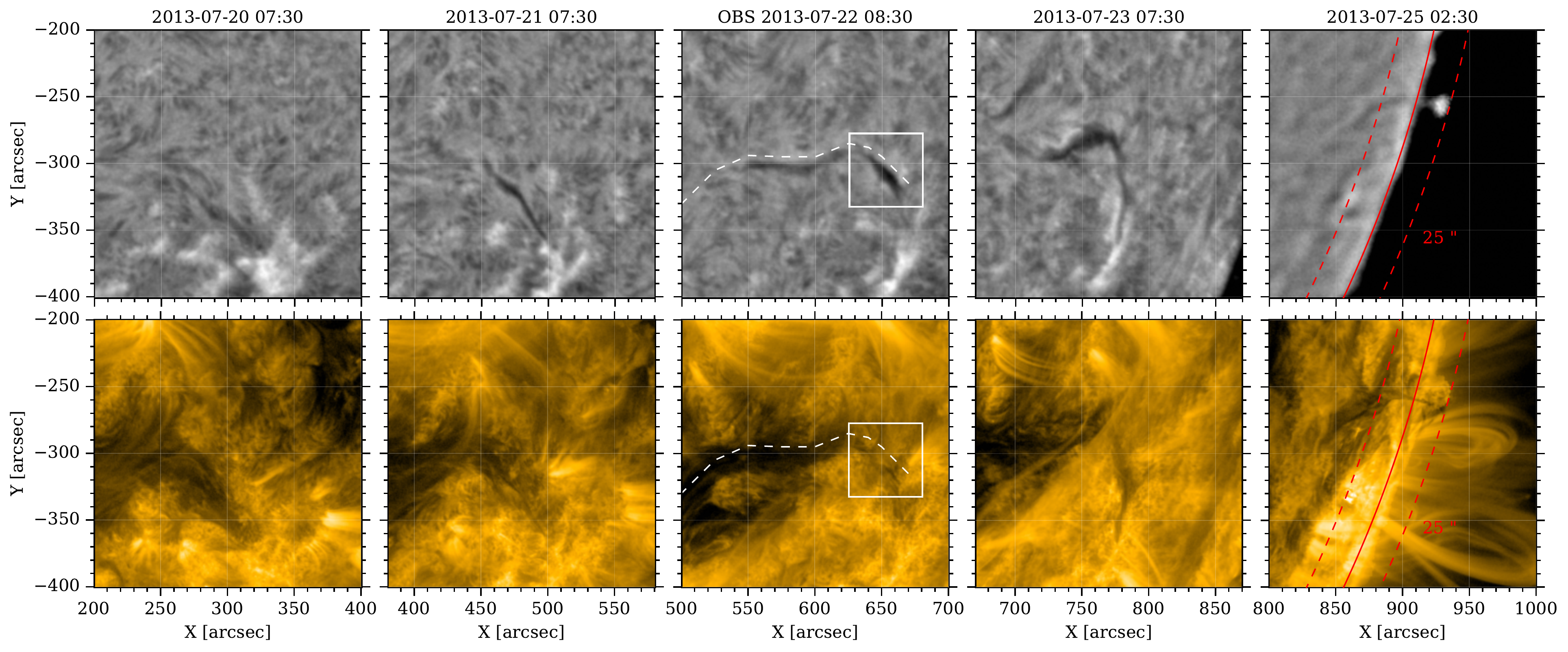}
\caption{The evolution of the filament during five days. We see how the filament is increasing its absorption and its size (up to 100\,\arcsec\ long) during its evolution. The top row shows images from GONG and the bottom row the same area but in the channel 171\,\AA\ of AIA. A dashed white line indicates the shape of the whole structure visually estimated from the combination of both filters.}
\label{fig:evolution}
\end{figure*}

Finally, to normalize the spectrum by the continuum intensity we used the Fourier Transform Spectrometer \citep[$I_{\mathrm{FTS}}(\lambda)$ or FTS,][]{neckel1984} atlas as a reference. Since the FTS is taken at disk center, we compare it with a spectrum taken at the same position (which we call $I_{0}$) a few hours before the observations that we want to analyze (which we call $I_{1}$), taken at other disk position (e.g., $\mu=0.65$). The average of a quiet-Sun region of the disk center observation is compared with the FTS atlas after convolving with the instrument PSF, $\sigma\sim$108\,m\AA:
\begin{equation}
\langle I_{QS0}\rangle \simeq  I_{C0} \cdot [I_{FTS}*G(\sigma)] \, .
\end{equation}

We then needed to compensate for changes in the observed intensities due to variations in the airmass during the observations, and take into account the ratio of the center-to-limb variation of the continuum intensity, i.e., $R_{\lambda\mu}=I_C(\lambda=8542.0, \mu=0.65) / I_C(\lambda=8542.0, \mu=1.0)$. This ratio is calculated from the values tabulated in \cite{Cox2000}. The new quiet-Sun continuum is:
\begin{equation}
I_{C1} = I_{C0} \cdot I_{WB1} / (I_{WB0}\cdot R_{\lambda\mu}) \, ,
\end{equation}
where $I_{WB}$ are the average intensity of the broadband images of each observation. After this, we only have to divide all the profiles by this quantity.

\subsection{Solar context}
\label{sec:solarContext}

To put the observations in context, we have made use of data provided by some of the band-pass filters \cite[AIA;][]{aia2012} on board NASA's \emph{Solar Dynamics Observatory} \cite[SDO;][]{sdo2012}, by the \emph{Global Oscillation Network Group} \citep[GONG;][]{hill1994} and from the \emph{Chromospheric Telescope} \citep[\CHROTEL;][]{Bethge2011}. We  display in Fig.~\ref{fig:sdomaps} images taken at a time in the middle of the observations (around 09:24~UT). The top left panel of Fig.~\ref{fig:sdomaps} shows an SDO/HMI image at the continuum of 6173\,\AA\ of the region where we can see the quiet Sun (QS) below our filament.

The rest of panels of Fig.~\ref{fig:sdomaps} show wavelengths sensitive to higher temperatures at chromospheric and coronal heights. Despite the lower resolution of these images ($\sim$1.2\,\arcsec/pix in HMI, $\sim$1.6\,\arcsec/pix in AIA, and $\sim$5\,\arcsec/pix in GONG) the filament is more visible in the filters sensitive to lower temperatures such as \ion{He}{ii} 304\,\AA, while less visible in the coronal filters \ion{Fe}{ix} 171\,\AA, \ion{Fe}{xii} 193\,\AA, and  \ion{Fe}{xiv} 211\,\AA. Other filters sensitive to larger temperatures (like \ion{Fe}{xviii} 94\,\AA) do not show any hint of the filament. Finally, the filament becomes clearly visible with a high absorption in the H$\alpha$ filter of GONG and the narrow filter of \ion{He}{i} of \CHROTEL\ (lower right panels).

Thanks to the temporal cadence of GONG and SDO, we can trace the evolution of the filament from its {formation} until it disappears behind the limb. Figure~\ref{fig:evolution} shows the evolution of the filament for five days. Throughout the evolution we see how the filament is increasing its absorption (it becomes darker) and its size. The top row shows the filament in H$\alpha$ taken from the GONG network, where the first four images are from GONG-Teide which has a better resolution, and the last one is from GONG-Mauna Loa. The second row shows the same area in the AIA/171\,\AA\ filter, but large coronal loops in the region mask the filament. Our spectropolarimetric observation was made two days after its formation (third column). Between the third and fourth day the filament {starts to decay} and it almost vanishes. A few hours later, the body starts to acquire absorption again. The last column shows the filament at the limb, where a small bright cloud appears to protrude by almost 25\,\arcsec\ above the limb. However, due to all the activity of the area, we cannot be sure that the filament seen in this frame is the same as that of our observation, as the physical properties of the region could have changed significantly in this period of time.

\section{Data analysis}
\label{sec:Dataanalysis}

After the complete reduction process, including fringes and crosstalk correction, we prepared the data to be analyzed with the \NICOLE\ code. To increase the signal-to-noise ratio of our observations, we average the entire time series since we do not see any substantial change in the evolution of the filament. Additionally, we also downsample the image with a binning of 4$\times$4 pixels, yielding a final sampling of 0.24\,\arcsec/pix. We clarify that all this processing is not really needed for Stokes $I$,  but it becomes necessary for the polarization to significantly improve the signal-to-noise ratio. We decrease the photon noise from $4-6\cdot10^{-3}$ to $2-6\cdot10^{-4}$ in units of the continuum intensity for Stokes $Q$, $U$, and $V$. Despite the high polarimetric sensitivity, some systematic artifacts are still present and we discuss them in the next sections.

\begin{figure*}[!ht]
\centering
\includegraphics[width=0.96\linewidth]{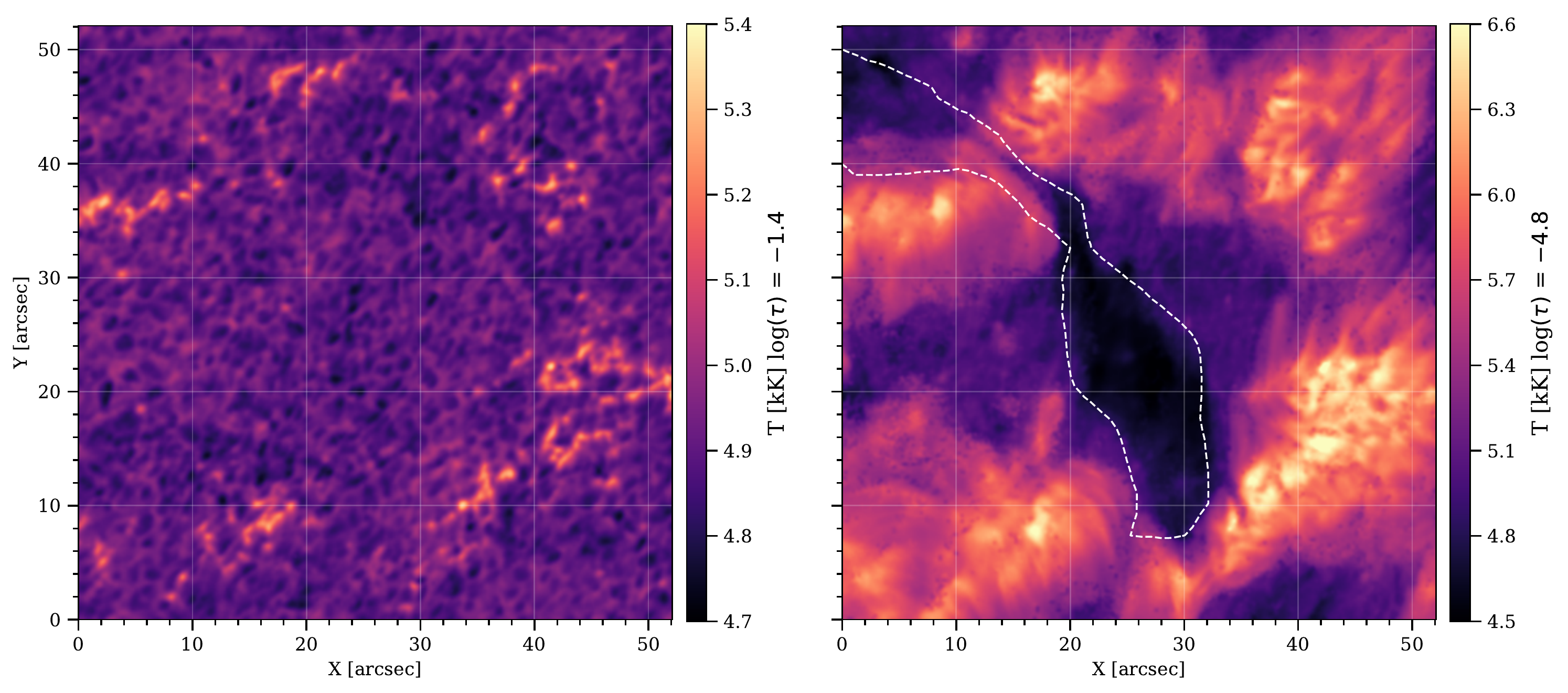}
\caption{Temperature stratification inferred with \NICOLE\ at $\log(\tau)=-1.4$ (left panel) and $\log(\tau)=-4.8$ (right panel). The white dashed contour draws the shape of the filament seen in H$\alpha$ by \CHROTEL.}
\label{fig:invTvB_temp}
\end{figure*}

\subsection{Inversion process}
\label{subsec:inversionProcess}

We use the \NICOLE\ inversion code \citep{socas2000,SocasNavarro2015} to infer the physical conditions {in the lower solar chromosphere}. Like other inversion codes  \citep[e.g. \Sir,][]{RuizCobo1992}, \NICOLE\ iteratively perturbs the physical parameters {of a model atmosphere} such as the temperature, line-of-sight velocity, and magnetic field to {find a synthetic spectra that can reproduce} the observations. The stratification of density and gas pressure are computed {by assuming} hydrostatic equilibrium (HE). {The physical parameters of the resulting model atmosphere are given as functions of the} optical depth scale at 5000\,\AA, $\log(\tau)$. {We have used a} \ion{Ca}{ii} model atom consisting of five bound levels plus continuum which is synthesized assuming {complete frequency redistribution (CRD) and including the effect of \ion{Ca}{ii}} isotopic splitting \citep{leenaarts2014}.

{We have used the spectral transmission profile of CRISP to degrade the emerging Stokes profiles during the inversion}. According to \cite{delaCruz2015}, the FWHM in this spectral range is $\sim$108\,m\AA\ and the Nyquist–Shannon (NS) sampling theorem establishes that the sampling has to be at least 54\,m\AA. As the observations were taken with a minimum sampling of 70\,m\AA, we synthesize the profiles with a wavelength axis with a sampling of 35\,m\AA. This new grid contains exactly all the observed points and fulfills the NS theorem. Moreover some points are added in the border of the wavelength range to avoid artifacts during the convolution. In a last step, our observed profiles are interpolated to the new grid but only the observed points are compared with the synthetic ones. We accomplish that by only weighting these points {accordingly} in the merit function.

\begin{figure*}[!ht]
\centering
\includegraphics[width=0.49\linewidth]{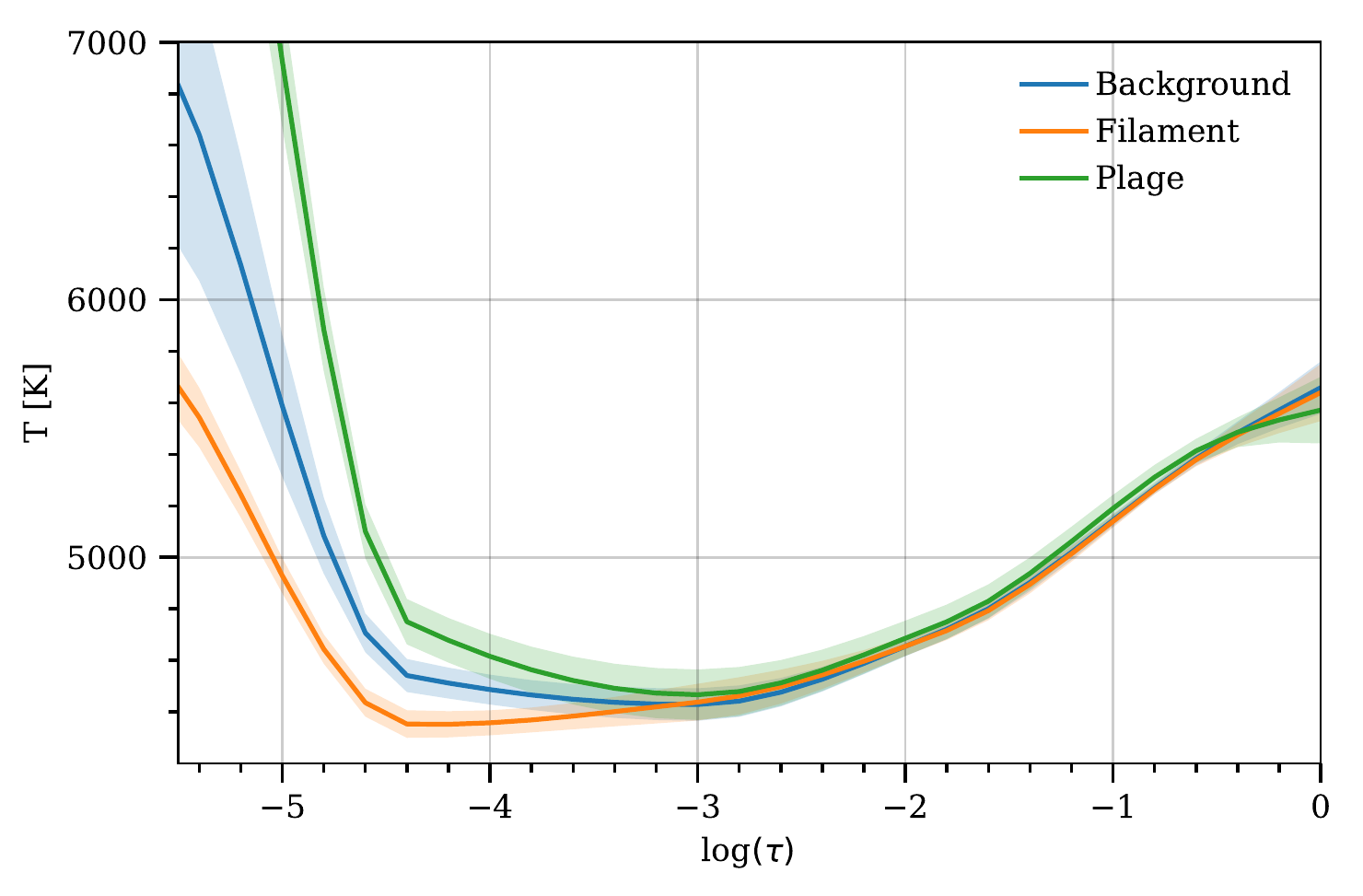}
\includegraphics[width=0.49\linewidth]{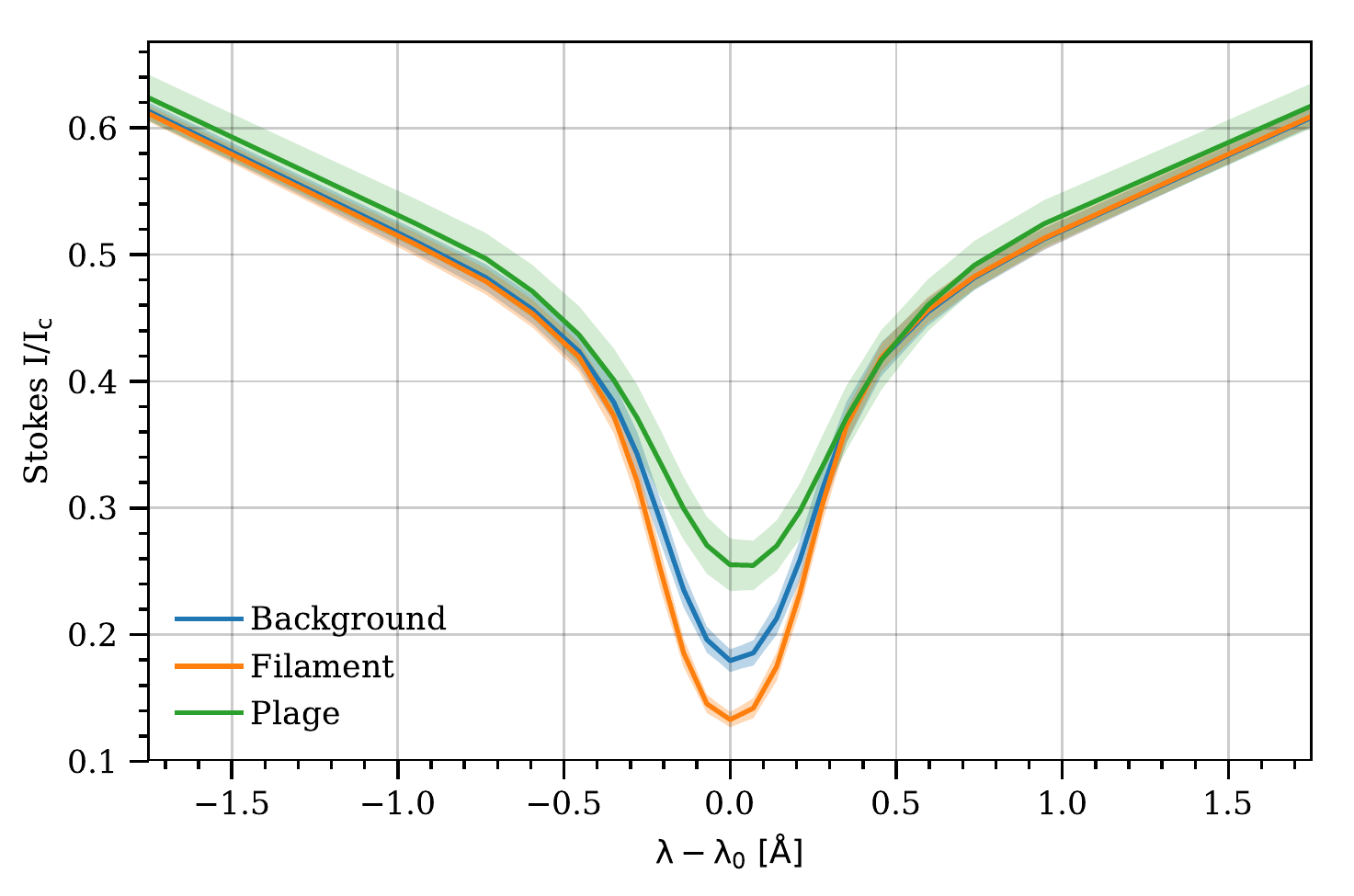}
\caption{Left panel: temperature stratification calculated from the average {of the temperature inferred from profiles of different regions}. Right panel: average Stokes $I$ profile {of the selected regions}. The coloured bands of each curve indicate the standard deviation of each set of profiles.}
\label{fig:invT}
\end{figure*}

In order to reduce the complexity of the problem and speed up the inversion, we decided to fix the microturbulent velocity. We have inverted several pixels across the FOV including different solar structures with the microturbulent velocity as a free parameter and we estimated an average value of 3\,\kms\ with a small dispersion of $\sim$1\,\kms. Microturbulence is often associated with line broadening produced by unresolved plasma motions. However, we want to stress that it can be slightly  degenerated to the temperature unless different spectral lines of different atoms are available. As we will see below, the average profiles of different regions (shown in the right panel of Fig.~\ref{fig:invT}) have a similar width, and their main difference is their depth. Although the absolute values of temperature could be slightly affected, {the main conclusions of the paper would remain unaffected}. We have also verified that the average Stokes profiles of the time series have the same broadening as that of individual profiles, meaning that the time scale of the filament evolution is larger than the duration of our series.

We have used the default equation of state (EOS) in \NICOLE\ to perform the full-FOV inversions. Concerning the upper boundary conditions for the gas pressure for the HE inversions we have used the same value of 0.3 dyn cm$^{-2}$ for all pixels. In the following, when pressure is mentioned, we are always referring to the total gas pressure. The inversions are done pixel by pixel, i.e., the NLTE problem is solved for each pixel independently assuming a plane-parallel atmosphere. This approximation is valid for the \ion{Ca}{ii}~8542\,\AA\ line where {horizontal photon scattering can be neglected} \citep{delaCruz2012}\footnote{See \cite{Stephan2016} for the limits of this assumption.}. To improve {the} convergence {of our inversions}, we performed three cycles. In the first cycle, we chose four nodes equidistantly located in the $\log(\tau)$ axis for temperature, one in line-of-sight (LOS) velocity and one more in each one of the three components of the magnetic field. This first cycle is very robust as there are no correlations between the magnitudes: the temperature describes the depth and width of the line, one node in LOS velocity mostly affects the core of the line (given the response function of the spectral line to the velocity and the dense sampling in the core), and the magnetic field describes the signal in Stokes $Q$, $U$, and $V$. Given the weak nature of the magnetic field {strengths that are involved} in our study, the \ion{Ca}{ii}~8542\,\AA\ line is in the weak field Zeeman regime and the Stokes $I$ is very weakly affected by the Zeeman broadening. The initial guess model is sampled with 43 points from $\log(\tau)$=--7.2 to $\log(\tau)$=+1.2. We start with a temperature stratification with an almost arbitrary parabolic shape as we do not need a better initial model to fit the spectral line. This robust first cycle is able to capture the main information and model the line profile with excellent convergence. To reduce possible artifacts, an horizontal smooth (in $\log(\tau)$ scale) is applied after each cycle. We use a median filter with a kernel size of 3 pixels to remove bad fits and a Gaussian filter with a FWHM of 3 pixels to smooth the whole FOV. This filter is applied at the same node positions of the inversion and then a Bezier-spline interpolation is applied \citep{delaCruz2013d}. The output of this smoothing process constitutes the input model for the next cycle. As strong asymmetries are not detected by an inspection of Stokes profiles and the FOV seems to be dynamically quite stable, only a maximum of two nodes are set to the LOS velocity in the second cycle. This is enough to capture the different velocities at the photospheric and chromospheric levels. A similar reasoning is applied to the LOS component of the magnetic field, as some Stokes $V$ profiles show several lobes indicating gradients of magnetic fields along the LOS. A last cycle is needed to improve the model of complex profiles located in the plage region. In this last cycle, the number of nodes in temperature is  increased to seven, i.e., we add one node between each pair of nodes from the previous cycle. The inversion strategy used to invert the full FOV is shown in table \ref{table:1}.

\begin{table}[!ht]
\caption{Nodes used to invert each pixel with the inversion code \NICOLE. Microturbulent and macroturbulent velocities are not inverted.}              
\label{table:1}      
\centering                                      
\begin{tabular}{c c c c c}          
\hline\hline                        
   Cycle & Temperature & v$_\mathrm{LOS}$ & B$_\mathrm{LOS}$ & B$_\perp$ \\
\hline                                   
   1 st &4 nodes &1 node &1 node &1 node\\
   2 nd &4 nodes &2 nodes &2 nodes &1 node\\
   3 rd &7 nodes &2 nodes &2 nodes &1 node\\
\hline                                             
\end{tabular}
\end{table}

\subsection{Temperature stratification}
\label{subsec:temperatureStratification}

To visualize the temperature stratification we have taken two horizontal slices at two different optical depths which are shown in Fig.~\ref{fig:invTvB_temp}: at $\log(\tau)=-1.4$, associated to a photospheric layer, and at $\log(\tau)=-4.8$, which provides a chromospheric view. These values have been chosen at two very sensitive layers {of the 8542 line in quiet Sun conditions} \citep{Quintero2016}. The left panel in Fig.~\ref{fig:invTvB_temp} {depicts} a very similar picture {to that in} the left panel in Fig.~\ref{fig:observation}, i.e., a photospheric {reversed} granulation {pattern} with temperatures of about 5000\,K. The second panel shows a cold filament with an average temperature of 4500\,K, surrounded by hot plage {regions} with a temperatures {larger} than 6500\,K.

To better study the {observed} area, we have classified the {spectral} profiles in three different classes according to the intensity in the core of the line. Filament profiles are those with intensities less than 0.148$I_c$\footnote{This number has been chosen by inspection to enclose the shape of the filament. The same applies to the plage threshold.}, plage profiles have intensities higher than 0.22$I_c$ and background profiles are those with intensities between the two. The left panel of Fig.~\ref{fig:invT} shows the average temperature stratification of each type of profile, the green line showing the temperature of the plage, the blue line displaying the temperature of the background and the orange line showing the temperature of the filament. We have extended the axis only to $\log(\tau)=-5.5$ because the line becomes insensitive after that point \citep{Quintero2016}. This figure demonstrates that the three curves overlap  below $\log(\tau)=-3$ and start to separate above. At $\log(\tau)=-5.5$ the difference between the filament and the environment is of around 1000\,K. The right panel of Fig.~\ref{fig:invT} displays the average profiles for each type. The shaded bands around the curves  indicate the standard deviation of each set of profiles (in both panels).

\begin{figure*}[!ht]
\centering
\includegraphics[width=0.96\linewidth]{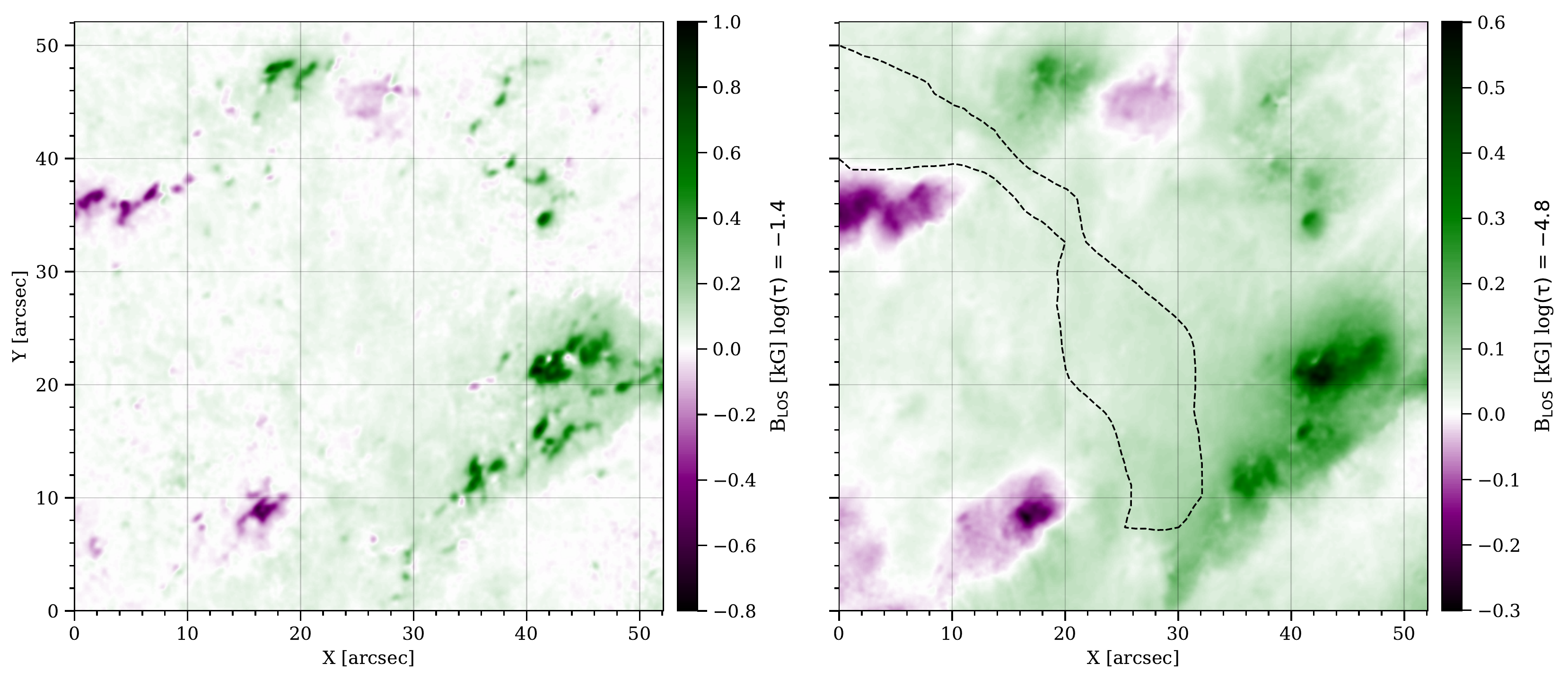}
\caption{Longitudinal component of the magnetic field inferred with \NICOLE\ at two levels: the $\log(\tau)=-1.4$ (left panel) and $\log(\tau)=-4.8$ (right panel).}
\label{fig:invTvB_blos}
\end{figure*}

\subsection{Longitudinal magnetic field}
\label{sec:LMF}

We focus now on the longitudinal component of the magnetic field as inferred by \NICOLE. We show this quantity at two heights in Fig.~\ref{fig:invTvB_blos}. As expected, the Stokes $V$ parameter encodes information from the photosphere to the chromosphere and the inversion process is able to extract  gradients of the LOS magnetic field with height.

In the photosphere (left panel of Fig.~\ref{fig:invTvB_blos}) the magnetic field appears more concentrated {at small scales}, while it is more diffuse in the chromosphere (right panel of the same figure), though maintaining the general topology of the photospheric field. The reason is that as the pressure decreases, the magnetic field expands. Probably, the magnetic field in the filament area appears uniform because we are tracing the upper part of the canopy, while the small loops are closing at lower heights. Although some small patches of opposite polarity can be seen in the photosphere (as well in the wings of Stokes $V$), we cannot ensure that these points really have a different polarity. They are usually close to the plage and hotter than the rest of the FOV. In such cases, two peaks at both sides of the core of the line usually appear in emission (in Stokes $I$) generating a Stokes $V$ signal with a different sign in the wings. This can be incorrectly interpreted by the inversion code as a reversal in the magnetic field. The FOV has a plage of almost 1\,kG in the photosphere that decreases up to 500\,G in the chromosphere, with a very weak magnetic field in the middle where the filament is located. One would expect to see the filament in these maps if the dense filament plasma is modifying the magnetic field lines of the region. However, there is no clear indication of its presence as we found similar values inside and outside the filament, unlike its clear detection in the temperature map. From this magnetic field map, it is hard to distinguish whether the filament is simply embedded in the lower atmosphere of the region with a similar magnetic field, or with a weaker magnetic field well above the solar surface and thus not detected.

\subsection{Transverse magnetic field}
\label{sec:TMF}

\begin{figure*}[!ht]
\centering
\includegraphics[width=\linewidth]{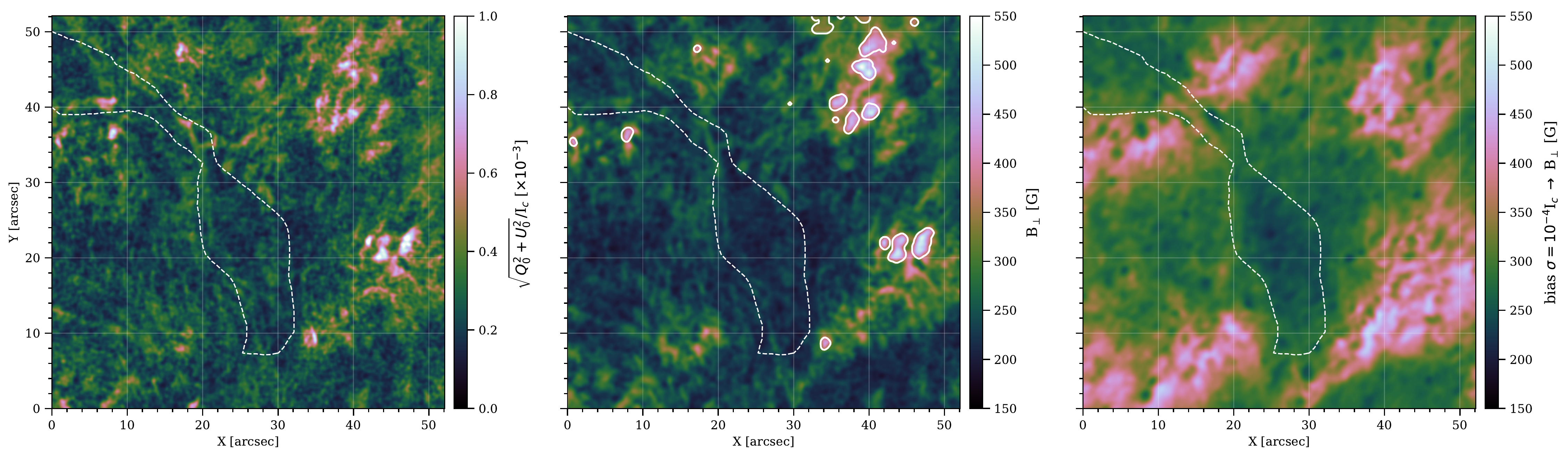}
\caption{Left panel: Linear polarization calculated at the core of the spectral line $\sqrt{Q(\lambda_0)^2+U(\lambda_0)^2}$. Middle panel: transversal magnetic field inferred during the inversion. Right panel: Magnetic field bias generated with a noise of $\sigma=10^{-4}\,I_c$. Some solid contours have been drawn in the left panel indicating where the inferred value is above the bias. The rest of values are overestimated. {The white dashed contour draws the shape of the filament seen in H$\alpha$ by \CHROTEL.}}
\label{fig:BLP}
\end{figure*}

We have used only one node to infer the transversal component of the magnetic field. It means that the inferred stratification is constant, with the same value at all the heights. The map of $B_\perp$ is shown in the left panel of the Fig.~\ref{fig:BLP}. This map shows higher values in the plage, while the filament is permeated by a very weak transverse magnetic field of less than 200\,G.

\begin{figure}[!ht]
\centering
\includegraphics[width=\linewidth]{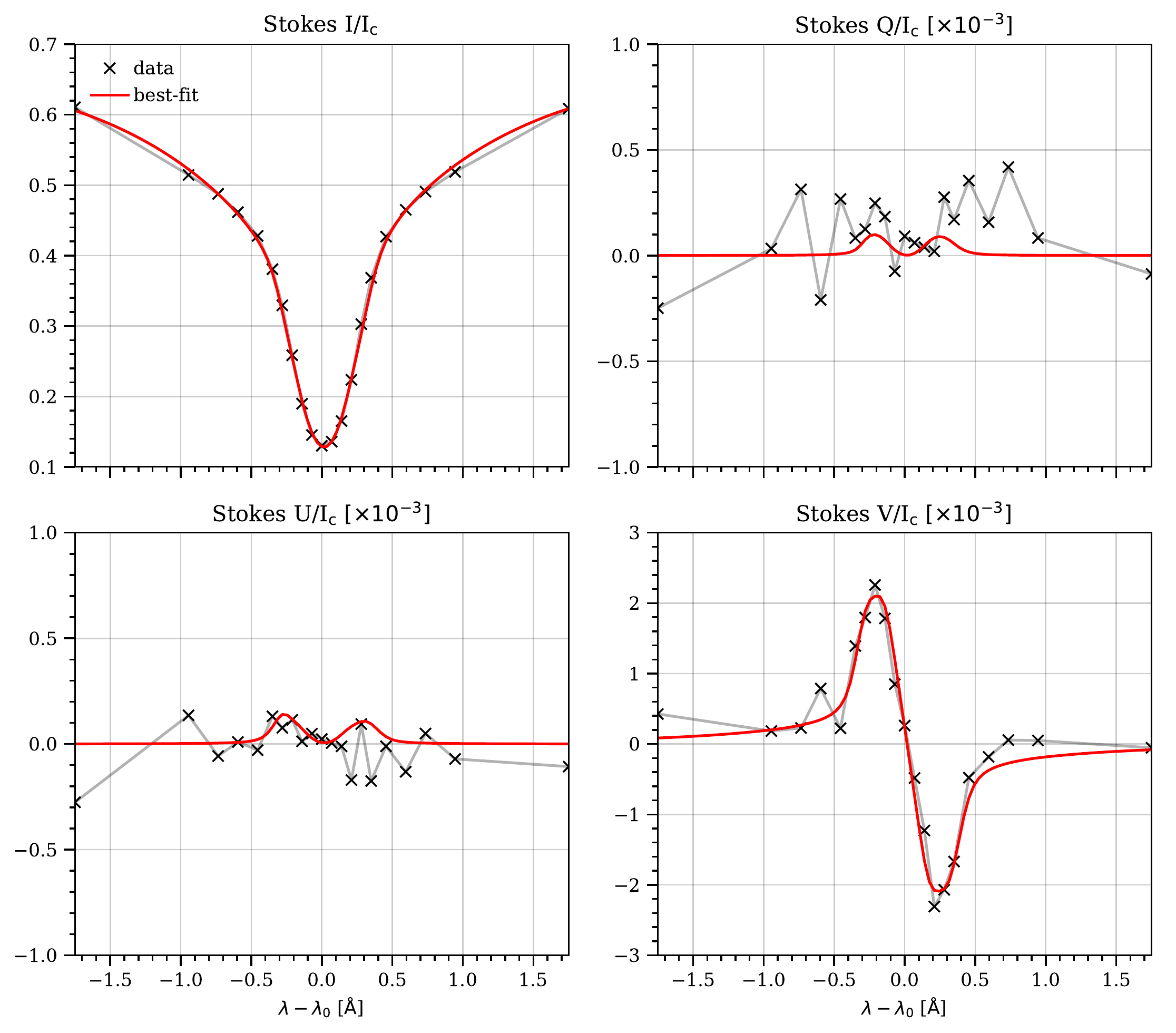}
\caption{A Stokes profiles sample from inside the filament (25\,\arcsec, 25\,\arcsec) shows how the fit was done properly, but as Stokes $Q/I_c$ and $U/I_c$ do not have enough signal, the inversion code can generate whatever profile with similar merit function values. The magnetic field inferred at $\log(\tau)=-5$ is $B_\parallel\sim$60\,G and $B_\perp\sim$150\,G.}
\label{fig:plot1}
\end{figure}

The linear polarization signals detected in the region are rather weak, of the order of $10^{-4}\,I_c$ in most of the map. {We show the line core linear polarization in the left panel of} Fig.~\ref{fig:BLP} {where only some very small patches show signals above $8\cdot10^{-4}\,I_c$}. We note that the residual systematic polarization artifacts that are present in our observations can strongly affect the observed Stokes profiles at the achieved S/N level. The latter are well above the photon noise and they do not decrease as we integrate a longer time series. An example of that effect is displayed in Fig.~\ref{fig:plot1}, which shows the Stokes profiles at position (25\,\arcsec, 25\,\arcsec) inside the filament. This figure shows that the fit is properly done. Stokes $Q$ and $U$ do not have enough signal, so that the inversion can fluctuate provided that the profiles produce similar merit function values. This poses some doubts on the {uncertainty} of the inferred transverse magnetic field.

\cite{marian2012} studied the effect of the noise in the estimation of the azimuth $\Phi_B$, inclination $\Theta_B$, longitudinal magnetic field $B_\parallel$, and transverse magnetic field $B_\perp$ in the weak field approximation. They found that the maximum likelihood estimation of the longitudinal component and the azimuthal angle are unbiased quantities. On the contrary, the {transverse} component, and hence the inclination angle, presents a non-zero bias, i.e., there is a difference between the value of the estimator and the true value of the parameter due to the presence of  noise. The value of the bias was derived by \cite{marian2012} and it is given by:
\begin{equation}
B_{\perp}^\mathrm{bias}(\sigma) = \left(\frac{-2\, \rm{ln}\,c}{\sum_j I_j''\,^2}\right)^{1/4}\frac{\sqrt{\sigma}}{C} \ ,
\label{eq:eq1}
\end{equation}
where $C=4.67.10^{-13}\lambda_0^2\sqrt{G}/2$, $G = 1.22$, and $\lambda_0=8542\,\AA$ in the case of this line. If the inferred $B_\perp$ is similar or smaller than the bias, this value may not be correct and may be strongly affected by the noise of the observations. Using the percentile 50 ($c=0.5$) and a very conservative value for the photon noise of $\sigma=10^{-4}\,I_c$, we obtain the map of bias displayed in the right panel of Fig.~\ref{fig:BLP}. The bias is proportional to the noise standard deviation so that, a lower value generates a lower bias and thus more reliable values of $B_\perp$.

We have drawn contours at the values of the bias in the left panel of Fig.~\ref{fig:BLP} indicating where the inferred value is above the bias. The contours show that there are only small patches where the inferred field is well above this bias and almost the rest of the map is below it. Therefore, the inferred value of $B_\perp$ is dominated by the noise in large part of the FOV, something that is expected given that Stokes $Q$ and $U$ have low polarization signals in the area. The spatial pattern of the bias, with low values in the center of the map and higher values in the border of the map coinciding with the plage, is a consequence of the line depth. The dependence of Eq.~\ref{eq:eq1} on the second derivative of the intensity profile produces a larger bias for shallower lines. Therefore, we can only give an upper limit on the transversal magnetic field of the filament of around 250\,G. Using the same formula, we can also estimate the noise required to have a bias below a certain magnetic field. For instance, for a magnetic field of $B_\perp=200$\,G, we estimate that one needs a photon noise below $\sigma=5\times10^{-5}\,I_c$ for a proper inference\footnote{This value is only estimated using the Zeeman effect, so if the Hanle effect is included in the calculation, this value will be slightly higher. {This value also assumes that the inference is calculated through the weak field approximation and therefore may differ slightly from the uncertainty calculated with NICOLE.}}.

\subsection{LOS velocity}
\label{sec:vlos8542}

\begin{figure*}[!ht]
\centering
\includegraphics[width=0.96\linewidth]{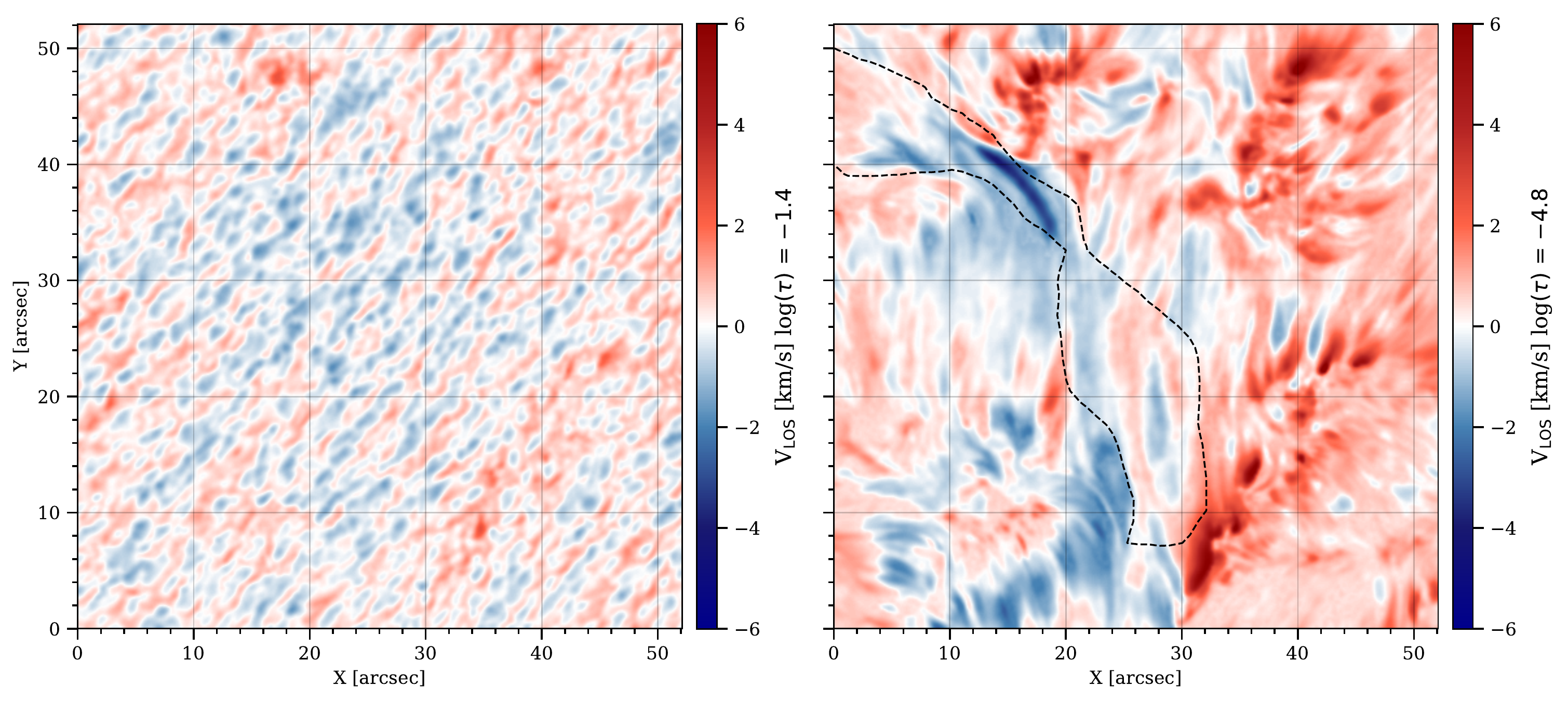}
\caption{Longitudinal velocity stratification inferred with \NICOLE\ at two atmospheric layers: $\log(\tau)=-1.4$ (left) and $\log(\tau)=-4.8$ (right). {The black dashed contour draws the shape of the filament seen in H$\alpha$ by \CHROTEL.}}
\label{fig:invTvB_vlos}
\end{figure*}

Figure~\ref{fig:invTvB_vlos} shows the LOS velocity at $\log(\tau)=-1.4$ and $\log(\tau)=-4.8$. After the inversion an offset was added to the whole velocity stratification to end up with the photosphere with zero mean velocity\footnote{Due to cavity errors (small imperfections of the reflecting surfaces of the etalons), the velocity map can show an artificial spatial gradient. To fix it, instead of adding a constant offset to the whole map, we have calculated a smooth version of the photospheric velocity map at $\log(\tau)=0$ by convolving the map with a Gaussian of a FWHM of 30 pixels, thus removing the fine  structure of the image. These values oscillate between $-0.8$\kms\ and $+0.2$\kms.}. The left panel shows small velocities across the FOV, with the typical granulation pattern with upflows in the granules and downflows in the intergranular lanes.

The right panel of Fig.~\ref{fig:invTvB_vlos} shows the velocity map at {the optical depth where we see the imprint of the filament also in temperature}. Fast downflows of almost 10\,\kms\ are found in the plage, while the area of the filament is almost at rest (with respect to the photosphere). A strong upflow of almost 5\,\kms\ is found at (15\,\arcsec, 40\,\arcsec), just where the "tail" of the filament is located.

\begin{figure}[!ht]
\centering
\includegraphics[width=0.96\linewidth]{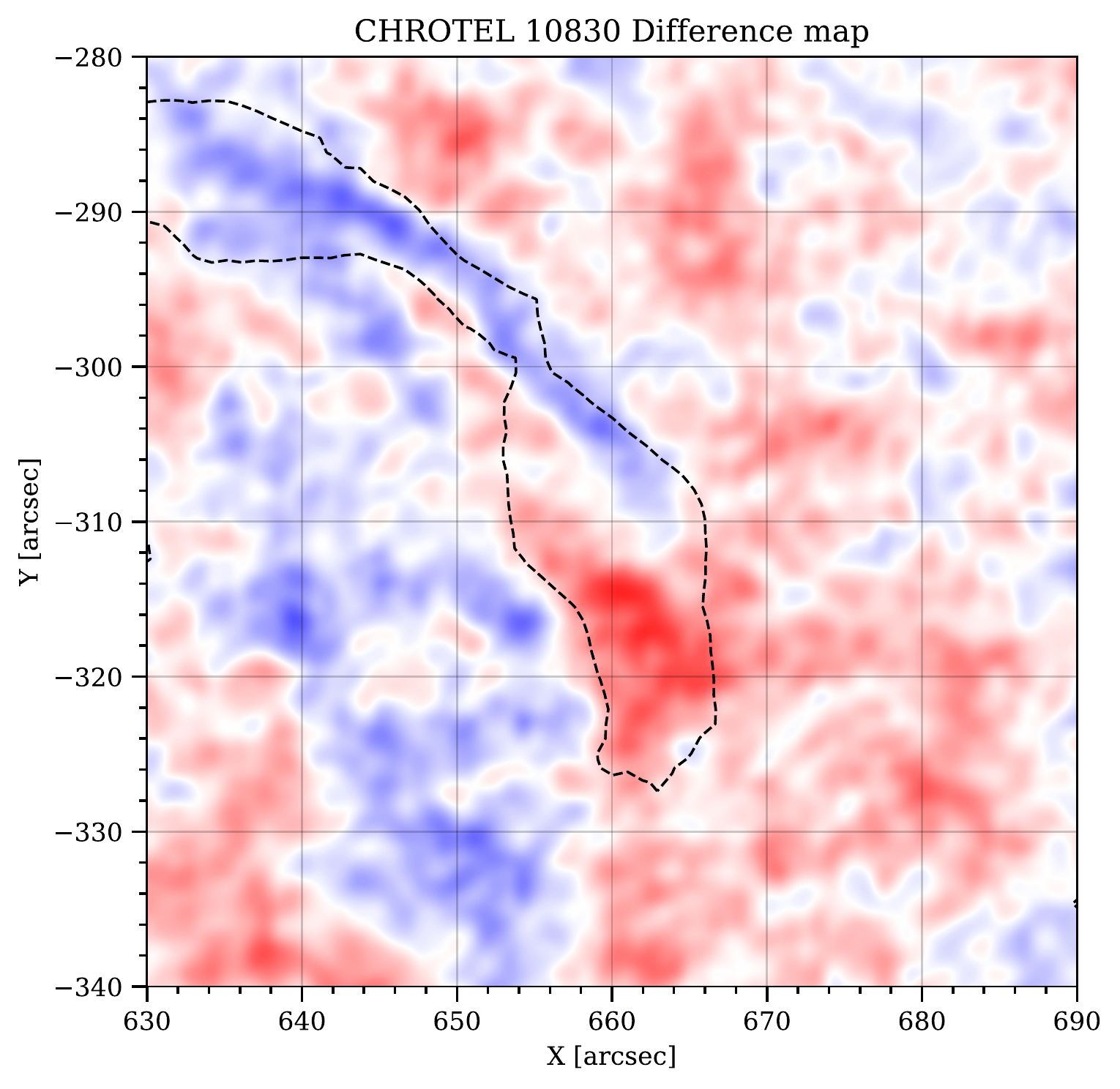}
\caption{Difference between the blue and red sides ($\pm 0.7$\,\AA) of the \ion{He}{i}~10830\,\AA\ with contours of H$\alpha$ taken by \CHROTEL. As in the previous figure, the blue represent motions towards the observer. The axes are in heliocentric coordinates.}
\label{fig:hemap}
\end{figure}

In order to have a better global picture and confirm the veracity of the velocity pattern of the filament, we have used observations from \CHROTEL. It is a robotic telescope for full-disk synoptic observations at the \ion{Ca}{ii}~K, H$\alpha$, and \ion{He}{i}~10830\,\AA\ chromospheric lines. \CHROTEL\ uses a tunable filter for observing the \ion{He}{i} line, which allows us to estimate the line-of-sight velocity. Due to the presence of several line components and the weak absorption of the multiplet, the calibration of the LOS velocity in \CHROTEL\ is difficult \citep{Bethge2011}. {For this reason, we only show {a Dopplergram computed from the subtraction of two} filtergrams at $\pm 0.7$\,\AA\ in arbitrary units (see Fig.~\ref{fig:hemap}). The velocity pattern detected in this line is very similar to that inferred for the \ion{Ca}{ii} line, displaying upflows of $\sim$7\,\kms\ in the border of the filament. Additionally, we also detected redshifted profiles with velocities of up to $\sim$5\,\kms\ in the lower part of the filament. These values have been calculated from some individual profiles. This might be a consequence of neutral material \citep{Khomenko2016} escaping from the structure because this is not detected in \ion{Ca}{ii}.}

We studied the surroundings of the filament to understand the flow of material. According to the middle panel of Fig.~\ref{fig:evolution} the filament seems to be the "tail" of a larger filament which covers the first half of the figure (indicated with a dashed white line) and could be feeding the filament with material. This flow of material could perfectly be parallel to the surface because it would have a component in the LOS direction due to the geometry.

\subsection{Transverse velocity}
\label{subsec:LCT}

A priori, it is not possible to diagnose transverse velocities with spectral lines because they produce no Doppler shift. We have used a Local Correlation Tracking algorithm \citep[LCT;][]{yi1995} to analyze the transverse flows in the images. This technique computes the intensity displacement between each two consecutive images and associates the optical flow to plasma velocities. We have used the Python implementation\footnote{The latest version is hosted at \url{https://github.com/Hypnus1803/FlowMapsGUI}} of this technique by \cite{campos2015}.

The applicability of this method to the chromosphere is very delicate. Changes in the intensity due to temperature or opacity changes might be interpreted by the code as something that is moving. In addition, inside the filament, where the contrast is very low, the estimation cannot be properly done. Finally, solar atmospheric changes can  affect the estimation of transverse velocities because they deform the solar images.

\begin{figure}[!ht]
\centering
\includegraphics[width=\linewidth]{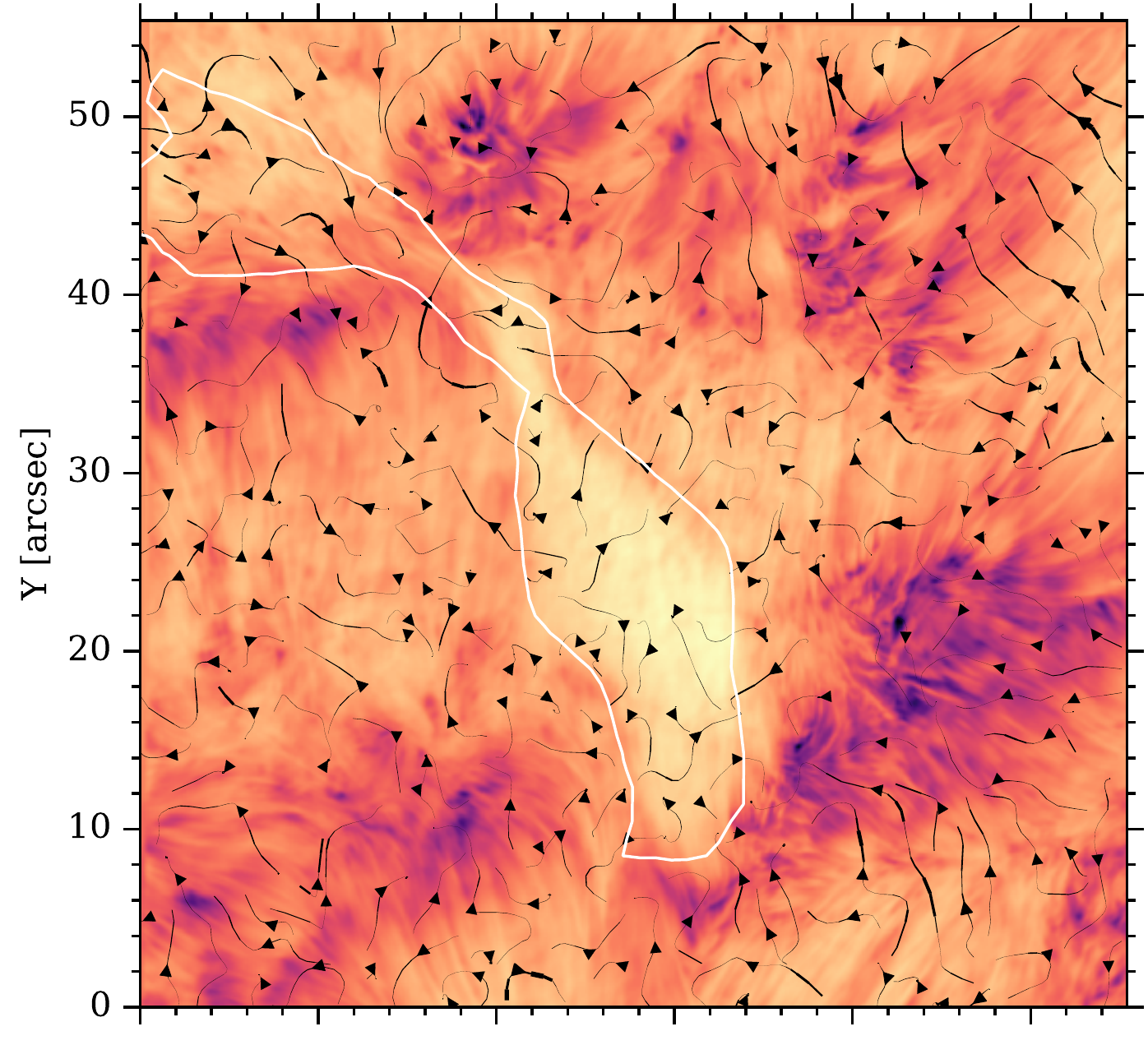}
\includegraphics[width=\linewidth]{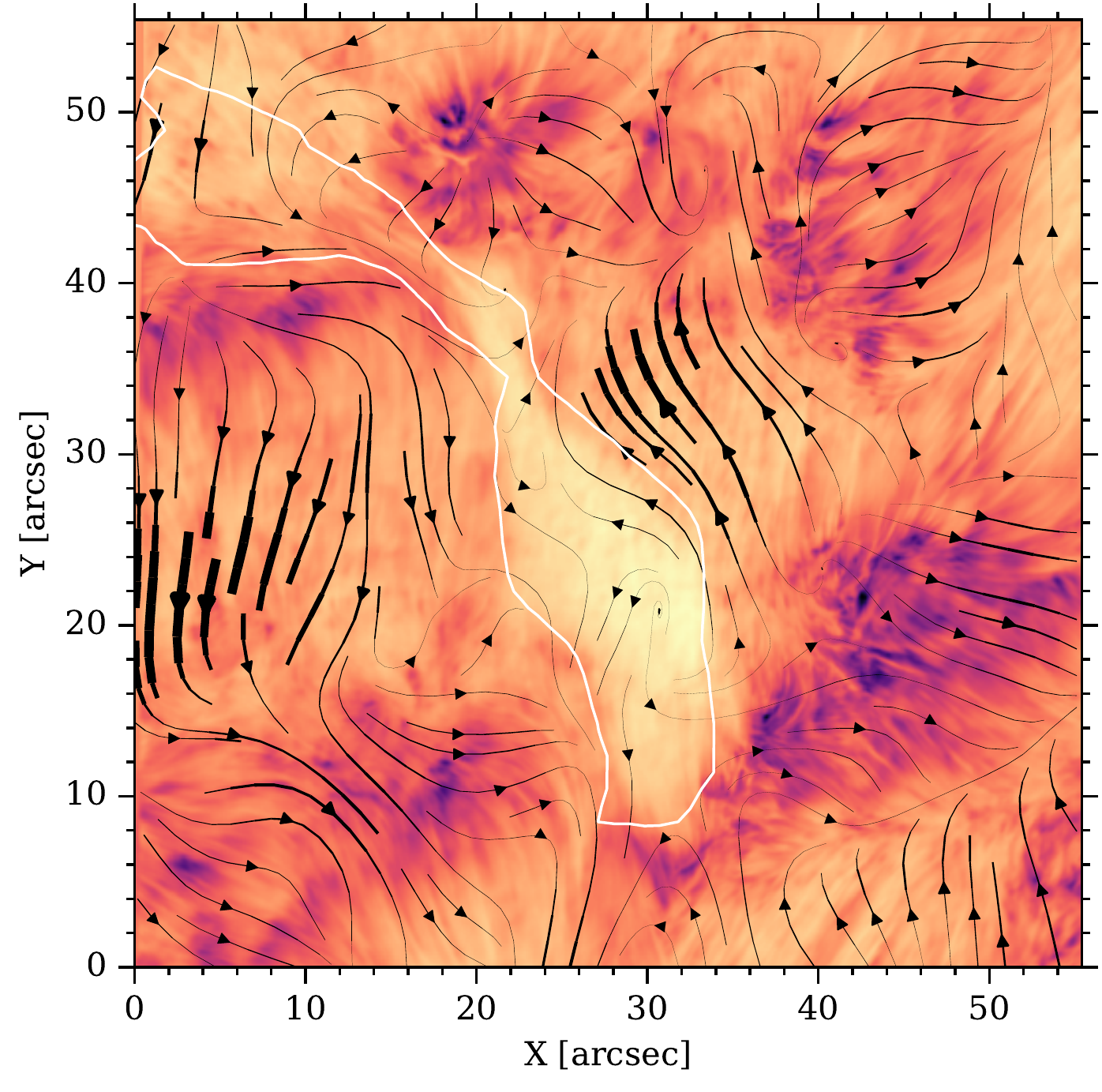}
\caption{Upper panel: map of the transverse component of the velocity inferred applying LCT to the temporal series at the wing of the \ion{Ca}{ii} line, tracing photospheric motions. Lower panel: The same quantity calculated in the core of the \ion{Ca}{ii} line, tracing chromospheric layers. The width of the arrows is proportional to the magnitude.}
\label{fig:vt}
\end{figure}

{Despite the aforementioned limitation of the technique, }we have applied the LCT method to the core images of the \ion{Ca}{ii} 8542\,\AA, where the filament is clearly visible and to the wing images at $-$1.8\,\AA\ from the line center. Since the filament is stable during the 14 minutes of the scan, we have taken the average of the flow estimated in all the frames. This allows us to cancel certain random motions or bad spurious estimations. We have used different spatial windows between 3\,\arcsec and 5\,\arcsec \citep{Diercke2018}, obtaining similar results. The main difference is that with a larger window the map is smoother and the values are a bit smaller.

Figure~\ref{fig:vt} shows the result of the average transversal velocity of the whole temporal series with the width of the arrows being proportional to the magnitude. The velocities at photospheric levels are around $\sim1$\,\kms\ while the velocity at chromospheric levels reach values of $\sim$2\,\kms. The background is the reverse image at the line core of the \ion{Ca}{ii} line, which helps to locate the position of the filament.

In the lower panel we detect a velocity field that seems to have the same direction as some fibrils seen in the center of the line, but the arrows are not always aligned with them. Another intriguing fact is that almost no transverse flow is found in the upper tail where we found high upflows of material, which can be explained, for instance, if the flow is constant and does not produce any change of brightness.

Another remarkable {feature} is given by the opposite motions on both sides of the filament. These horizontal {motions} reach {systematic} mean flow speeds of $\sim$1.5\,\kms (averaged  over 14 minutes). While other studies such as \cite{Zirker1998} or \cite{Diercke2018} find counterflows inside the filament and they associate these dynamics with internal motions of the filament itself, our counter-motions are detected mainly outside the filament \cite[like in the case of ][]{yi1995} and could be associated, for instance, with the formation of the filament axis through photospheric shearing motions \citep{van1989}. To analyze whether there are shearing motions in the photosphere, we have applied the LCT method also to the photosphere using the far wings of the \ion{Ca}{ii} line. This result is displayed in the upper panel of Fig.~\ref{fig:vt} and it does not show any pattern that indicate these motions, but a  chaotic granular pattern. A possible explanation for this non-detection could be that such motions have stopped or that our temporal range is not enough to recover these possible small velocities (perhaps below the granulation velocities).

We detect that the flows in the bright plage zones are outward, something counter-intuitive since from the \NICOLE\ inversion the material seems to be falling at those points. One possible explanation for this effect could be that the LCT method is also sensitive to changes in brightness produced by wave propagation. Waves moving away from areas of high magnetic field and running along field lines would produce such artifacts \citep{Bloomfield2007}. This poses a question whether the opposite motions found in the filament by the LCT method may not be real, but rather changes in the intensity, indicating that the LCT method should not be safely applied to the chromosphere. Therefore, to differentiate what is actually moving and what are propagating waves we would need to make a more detailed study of the temporal evolution of these oscillations at different heights \citep{Vecchio2007} with a larger time series since 14 minutes is not enough to study them through a Fourier analysis.

\section{Supporting the filament plasma}
\label{sec:support}

\begin{figure*}[!ht]
\centering
\includegraphics[width=0.48\linewidth]{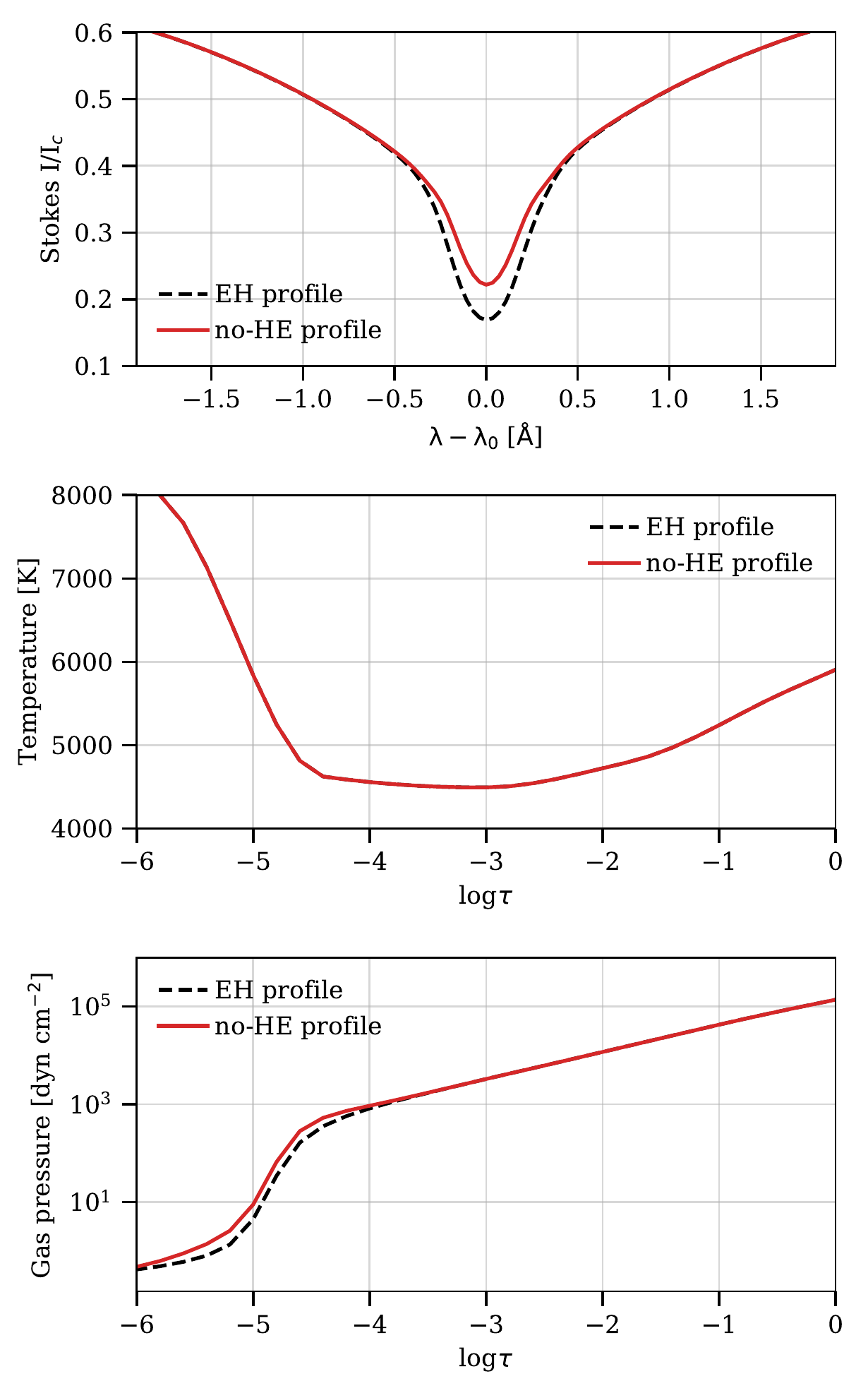}
\includegraphics[width=0.48\linewidth]{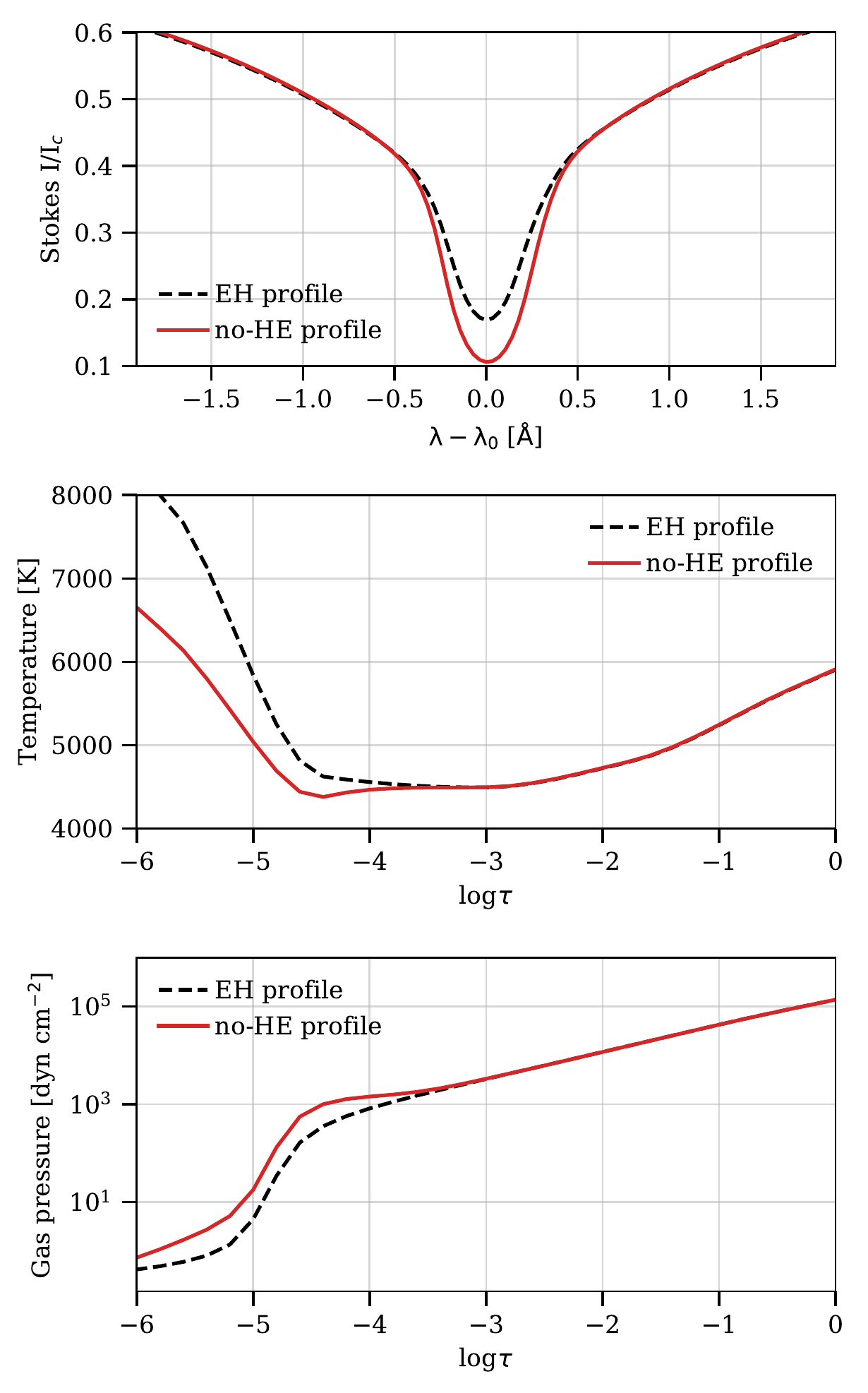}
\caption{Stokes profiles synthesized under no-HE from the modifications of the model of QS under HE. Left column: only when the pressure (or density) is increased. Right column: when the temperature is decreased (even with an overdensity).}
\label{fig:plotRIA}
\end{figure*}

\begin{figure*}[!ht]
\centering
\includegraphics[width=0.96\linewidth]{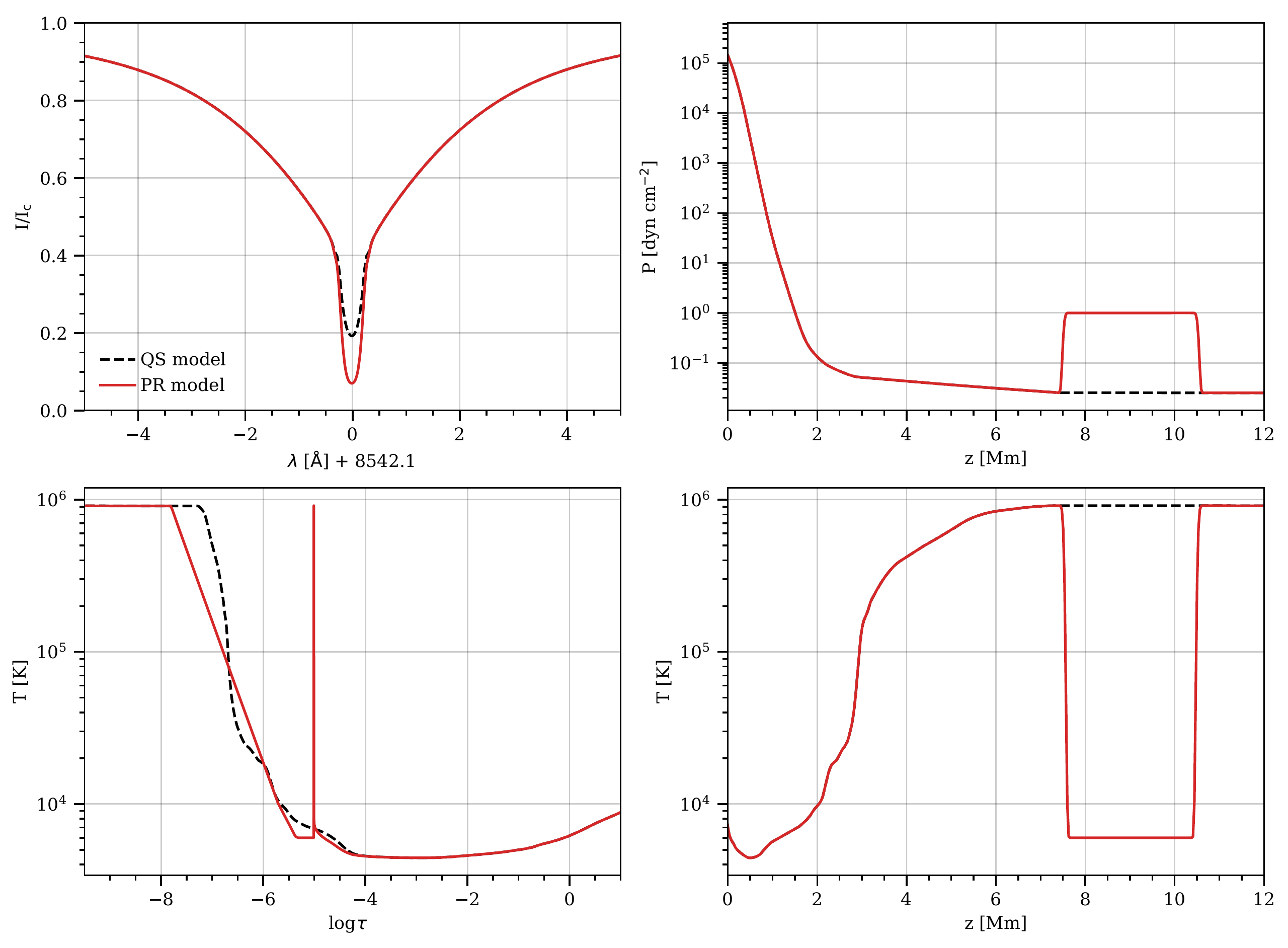}
\caption{Upper left panel: Stokes $I$ profile generated from  two different atmospheres, a QS model (dashed black line) and a prominence above it (red solid line). Upper right panel: gas pressure in geometrical height scale. Lower left panel: temperature in optical depth scale. Lower right panel: temperature in geometrical height scale.}
\label{fig:prominence}
\end{figure*}

The general conclusion of the inversions is that we have found a filament with a lower temperature than its environment (approximately 1000\,K colder) and a higher density (about 20 times larger). These values may vary slightly depending on the height at which we analyzed the results. Specifically, these values refer to $\log(\tau)\sim-5$. These results are consistent with the observational definition of these structures: {\it a filament is cool chromospheric plasma overdensity embedded in the hot and low dense solar corona} \citep{mackay2010}.

With these properties, the filament will persist at chromospheric/coronal heights because of the magnetic field that supports the plasma against the gravitational force. The magnetic field (and in particular its topology) seems to be the main ingredient that allows this sustentation and multiple theoretical models have been proposed to explain the presence of this material for so long \citep{Antiochos1994,van1989,Martens2001,vanB2010,Luna2015}.

However, we note that the assumptions made for the inversions are not fully compatible with the description above. \NICOLE\ assumes that the atmospheric models are in hydrostatic equilibrium. This implies that the gas pressure is computed from the temperature {optical depth scale}, neglecting the Lorentz (magnetic) force by solving, in a plane-parallel atmosphere:
\begin{equation}
\frac{dP(z)}{dz} = -\rho(z)g_\odot(z)  \ \overset{\mathrm{ideal\, gas}}{\rightarrow }\ \frac{dP(z)}{dz} = -\frac{P(z)}{T(z)} \left(\frac{\mu m_H g_\odot}{k_B}\right) \ ,
\label{eq:hidrostatic}  
\end{equation}
where $g_\odot$ is the solar surface acceleration, $k_B$ the Boltzmann's constant, $\mu$ the mean molecular weight and $m_H$ the mass of an hydrogen atom. The equilibrium is obtained between the pressure and gravity forces. A consequence of the previous equation is that the chromospheric material under this approximation is not supported by the magnetic field and it is only the gas pressure gradient that compensates the gravitational force.

An obvious question is then whether the filament is really supported against gravity by the hydrostatic equilibrium or not. This is a reasonable question given that the model is capable of reproducing the spectral profile of the \ion{Ca}{ii} line. To understand how \NICOLE\ is able to find such a solution it is more convenient to discuss it in terms of geometric height. If we transform the stratification from optical depth to geometric height, we would have our filament at a height of 1500\,km. At this height, the pressure can still support a material with these properties. However if this overdensity is at higher heights where pressure has decreased some orders of magnitude, this material cannot be supported. The interface between these two layers will become unstable under any disturbance and a Rayleigh-Taylor instability can quickly develop and dilute the body of the prominence \citep{Khomenko2014RTI}. In this case, the presence of a magnetic field is crucial to support this plasma and damp the formation of such instability.

Concerning this magnetic field, based on the inference from the spectropolarimetric data, the longitudinal component has values around 60\,G (see Sec.~\ref{sec:LMF}), while we can only give a maximum value of 250\,G to the transverse component due to the noise level in Stokes $Q$ and $U$ (see Sec.~\ref{sec:TMF}). This gives us a maximum value for the magnetic field strength of the order of $\sqrt{B_\parallel^2+B_\perp^2}\sim$260\,G, a result still quite high compared to the values found in the literature \citep{Gunar2016}.

Checking if the filament is in hydrostatic or magnetic equilibrium from observations is a difficult task. To shed some light, we have carried out  several experiments. In the first numerical experiment, we start from the atmospheric model in HE inferred for the QS and analyze the influence on the emergent profile of modifications in the temperature and the pressure. To this end, we synthesize the profiles with \NICOLE\ without imposing HE and adding some perturbations. The results are shown in Fig.~\ref{fig:plotRIA}, where the original stratification is shown as a black dashed line. Since the new material will not be in hydrostatic equilibrium and we want to study only the thermodynamical properties, we assume that a magnetic field is always acting to keep the material static, compensating the extra perturbations introduced by us.

The left panel of Fig.~\ref{fig:plotRIA} shows an atmosphere where the pressure has been increased at the chromosphere around $\log(\tau)=-5$ (equivalent to a density increase). This scenario shows a dense filament model over a QS atmosphere. The result of the synthesis shows a Stokes $I$ profile with less absorption, drawn as a solid red line. The increase in pressure (or density) causes an increase in the source function, mainly due to the  higher increase of the upper level population, relatively to the increase of the lower level population through the radiative rates.

On the other hand, the right panel of Fig.~\ref{fig:plotRIA} shows an atmosphere where we have decreased the temperature also in the chromosphere around $\log(\tau)=-5$. In this case, the synthesis gives a deeper profile, similar to the one found inside the filament. As the sensitivity of the line to changes in temperature is higher than to changes in gas pressure, one could still increase the pressure (as one would expect for an overdense filament) and the line profile would be still deeper than that of the QS. The resulting stratification is very similar to the solution found in HE. This first experiment shows that an overdensity must be colder than the environment to reproduce the deepest profiles observed. But given the degeneration that exists between temperature and density (although the profile is more sensitive to changes in temperature), we can continue generating the same profile by increasing the density and decreasing the temperature. This does not allow us to find a unique solution.

Now, we investigate the effect of the height of the filament on the formation of the spectral line. To do this, we simulate the scenario of a filament of similar characteristics located at coronal heights. Because \NICOLE\ has not been developed to correctly address the external part of the atmosphere where high temperatures are present (transition region and corona), we have used the RH code \citep{Uitenbroek2001} which includes the ingredients for that purpose.

In order to have an extended corona where to place the filament, we assume that the filament is embedded in a QS atmosphere that we have taken from a snapshot from a 3D MHD simulation calculated with the \BIFROST\ code \citep{Gudiksen2011}. This simulation \citep{Carlsson2016} is publicly available as part of the Interface Region Imaging Spectro-graph (IRIS) mission \citep{IRIS2014}. To compute the QS model, we horizontally average the snapshot 541 (similar to the used by \citealt{Quintero2016} and \citealt{delaCruz2013h}). As the simulation provides height ($z$), gas pressure ($P$), and temperature ($T$), we transform them into the quantities required by the RH code, which are: mass column density ($\rho$), temperature ($T$), electron density ($N_e$), LOS velocity ($v_{los}$), and microturbulent velocity ($v_{turb}$) (for more details see \citealt{Uitenbroek2001}). The synthesis from the original model is shown in Fig.~\ref{fig:prominence} in dashed black line. On top of this QS model we placed a filament of typical properties \citep{Tziotziou2007,mackay2010,Panesar2014} with a temperature of 8000\,K, pressure of 1\dyn, and vertical size of 3\,Mm at a height of 9\,Mm. The resulting Stokes $I$ profile is shown in red in the same figure.

The model shown in Fig.~\ref{fig:prominence} has a pressure profile that decreases rapidly to the chromosphere (2\,Mm) and then more slowly. The temperature profile includes a smooth transition region, wider than the individual stratifications of the model due to the horizontal averaging, finally reaching a corona at $10^6$\,K. The upper left panel shows the synthesis of Stokes $I$ from both models: the QS model and the filament model. This  demonstrates that a deeper Stokes profile (similar to the one observed) can be generated with a filament at coronal heights under no hydrostatic equilibrium conditions.

On the other hand, another important conclusion can be drawn in the transformation from geometric scale to optical depth scale. The lower left panel of Fig.~\ref{fig:prominence} shows that the filament, physically located at coronal heights, is at the same optical depths as the original chromosphere, i.e., $\log(\tau)\sim-5$. The reason of that is the small opacity of the material between the filament and the upper chromosphere that contributes almost nothing to the optical depth. This is an indication that the line is really sensitive to the filament and its conditions, independently of its height. The latter is reasonable because the filament is clearly visible in the core of the line.

Summarizing, the \ion{Ca}{ii}~8542\,\AA\ spectral line is sensitive to the atmospheric conditions of the filament. However, due to the degeneration that exists among height, temperature, and pressure, we cannot give an accurate estimation of these three parameters. Other chromospheric lines with different sensitivity to temperature and pressure, such as \ion{Ca}{i} 4227\,\AA, may be suitable for restricting the possibilities. To infer the height we should observe the filament from another perspective (e.g. in the limb) or by triangulation  \citep[e.g., using STEREO,][]{orozco2014,schad2016}.

\section{Summary and conclusions}

We have used the \NICOLE\ code to study a solar filament observed on the \ion{Ca}{ii}~8542\,\AA\ line. Through a careful data reduction and inversion we have studied the stratification of temperature, density and magnetic field of a region containing a filament. Thanks to the high sensitivity of the line we have inferred the atmospheric conditions from the photosphere to the chromosphere, i.e., $\log(\tau)\sim[\,0,-5.5\,]$. From this inference we have found that the filament appears as an overdensity (20 times greater than the environment) and with colder temperatures (about 1000\,K lower than the environment). These results are in concordance with existing literature.

However, \NICOLE\ assumes hydrostatic equilibrium during the inversion process. If this condition is not fulfilled in the Sun, we have shown that the degeneracy  between  temperature and pressure (or density) in the formation of the intensity profile makes it impossible to estimate these values accurately. We have also shown that we are insensitive to the actual height of the filament, as the small opacity of the material between the filament and the upper chromosphere contributes almost nothing to the optical depth. Therefore, even being sensitive to the "local" filament conditions, it is not possible to use only this spectral line to distinguish between an hydrostatically supported filament at chromospheric heights and a colder and denser filament at coronal heights.

Concerning the magnetic field, we have detected homogeneous fields with a strength smaller than 260\,G throughout the region without a clear distinct value on of the filament. Weaker fields exist in the area but we are not able to detect them due to the noise level in the observations. These values indicate that we are in a field regime in which the Hanle effect begins to play an important role and should be taken into account \citep{manso2010}.

Although the filament is stable during the observations, {our observations reveal gas motions operating at very small scales}. From a global point of view, our filament appears to be attached to a larger filament, through which material could be transferred. We detect it as a blueshifted flow in the tail of the observed filament. From a local point of view, we have detected, using the \ion{He}{i}~10830\,\AA\ line, higher velocities of material falling from the filament than those inferred from the \ion{Ca}{ii} line. This material loss due to the neutral nature of \ion{He}{i} (not sensitive to the magnetic field) is expected. These differences in velocity reminds us that the filament is not only supported by the gas pressure.

Finally, we have also studied the time series using LCT. As a result, we obtain flows practically aligned with the fibrils but with directions opposite to the velocities obtained by the Doppler-shift of the line. One possible explanation would be the propagation of waves emerging from the magnetic field concentrations of the plage and could serve as a diagnosis of the field topology. However, more work remains to be done to confirm that.

\begin{acknowledgements}
We would like to thank the anonymous referee for its comments and suggestions.
Financial support by the Spanish Ministry of Economy and Competitiveness 
through projects AYA2014-60476-P and AYA2014-60833-P are gratefully acknowledged. CJDB acknowledges Fundaci\'on La Caixa for the financial support received in the form of a PhD contract. AAR also acknowledges financial support through the Ram\'on y Cajal fellowships. This research has made use of NASA's Astrophysics Data System Bibliographic Services. 
JdlCR is supported by grants from the Swedish Research Council (2015-03994), the Swedish National Space Board (128/15) and the Swedish Civil Contingencies Agency (MSB). This project has received funding from the European Research Council (ERC) under the European Union's Horizon 2020 research and innovation programme (SUNMAG, grant agreement 759548).
The Swedish 1-m Solar Telescope is operated on the island of La Palma by the Institute for Solar Physics of Stockholm University in the Spanish Observatorio del Roque de los Muchachos of the Instituto de Astrof\'isica de Canarias. The Institute for Solar Physics is supported by a grant for research infrastructures of national importance from the Swedish Research Council (registration number 2017-00625). This research has made use of NASA’s Astrophysics Data System Bibliographic Services.

We acknowledge the community effort devoted to the development
of the following open-source packages that were used in this work: numpy
(numpy.org), matplotlib (matplotlib.org)

This research has made use of SunPy, an open-source and free community-developed solar data analysis package written in Python \citep{sunpy2015}.

The author thankfully acknowledges the technical expertise and assistance provided by the Spanish Supercomputing Network (Red Espa\~nola de Supercomputaci\'on), as well as the computer resources used: the LaPalma Supercomputer, located at the Instituto de Astrof\'isica de Canarias.

\end{acknowledgements}


\begin{thebibliography}{}
\makeatletter
\relax
\def\mn@urlcharsother{\let\do\@makeother \do\$\do\&\do\#\do\^\do\_\do\%\do\~}
\def\mn@doi{\begingroup\mn@urlcharsother \@ifnextchar [ {\mn@doi@}
  {\mn@doi@[]}}
\def\mn@doi@[#1]#2{\def\@tempa{#1}\ifx\@tempa\@empty \href
  {http://dx.doi.org/#2} {doi:#2}\else \href {http://dx.doi.org/#2} {#1}\fi
  \endgroup}
\def\mn@eprint#1#2{\mn@eprint@#1:#2::\@nil}
\def\mn@eprint@arXiv#1{\href {http://arxiv.org/abs/#1} {{\tt arXiv:#1}}}
\def\mn@eprint@dblp#1{\href {http://dblp.uni-trier.de/rec/bibtex/#1.xml}
  {dblp:#1}}
\def\mn@eprint@#1:#2:#3:#4\@nil{\def\@tempa {#1}\def\@tempb {#2}\def\@tempc
  {#3}\ifx \@tempc \@empty \let \@tempc \@tempb \let \@tempb \@tempa \fi \ifx
  \@tempb \@empty \def\@tempb {arXiv}\fi \@ifundefined
  {mn@eprint@\@tempb}{\@tempb:\@tempc}{\expandafter \expandafter \csname
  mn@eprint@\@tempb\endcsname \expandafter{\@tempc}}}

\bibitem[\protect\citeauthoryear{{Antiochos}, {Dahlburg}  \&
  {Klimchuk}}{{Antiochos} et~al.}{1994}]{Antiochos1994}
{Antiochos} S.~K.,  {Dahlburg} R.~B.,   {Klimchuk} J.~A.,  1994, \mn@doi
  [\apjl] {10.1086/187158}, \href
  {http://adsabs.harvard.edu/abs/1994ApJ...420L..41A} {420, L41}

\bibitem[\protect\citeauthoryear{{Asensio Ramos}, {de la Cruz Rodr\'{\i}guez},
  {Mart\'{\i}nez Gonz\'{a}lez}  \& {Socas-Navarro}}{{Asensio Ramos}
  et~al.}{2017}]{asensio2017}
{Asensio Ramos} A.,  {de la Cruz Rodr\'{\i}guez} J.,  {Mart\'{\i}nez
  Gonz\'{a}lez} M.~J.,   {Socas-Navarro} H.,  2017, \mn@doi [\aap]
  {10.1051/0004-6361/201629755}, \href
  {http://adsabs.harvard.edu/abs/2017A\%26A...599A.133A} {599, A133}

\bibitem[\protect\citeauthoryear{{Aulanier} \& {D\'{e}moulin}}{{Aulanier} \&
  {D\'{e}moulin}}{2003}]{Aulanier2003}
{Aulanier} G.,  {D\'{e}moulin} P.,  2003, \mn@doi [\aap]
  {10.1051/0004-6361:20030227}, \href
  {http://adsabs.harvard.edu/abs/2003A\%26A...402..769A} {402, 769}

\bibitem[\protect\citeauthoryear{{Bethge}, {Peter}, {Kentischer},
  {Halbgewachs}, {Elmore}  \& {Beck}}{{Bethge} et~al.}{2011}]{Bethge2011}
{Bethge} C.,  {Peter} H.,  {Kentischer} T.~J.,  {Halbgewachs} C.,  {Elmore}
  D.~F.,   {Beck} C.,  2011, \mn@doi [\aap] {10.1051/0004-6361/201117456},
  \href {http://adsabs.harvard.edu/abs/2011A\%26A...534A.105B} {534, A105}

\bibitem[\protect\citeauthoryear{{Bloomfield}, {Lagg}  \&
  {Solanki}}{{Bloomfield} et~al.}{2007}]{Bloomfield2007}
{Bloomfield} D.~S.,  {Lagg} A.,   {Solanki} S.~K.,  2007, \mn@doi [\apj]
  {10.1086/523266}, \href {http://adsabs.harvard.edu/abs/2007ApJ...671.1005B}
  {671, 1005}

\bibitem[\protect\citeauthoryear{{Campos Rozo} \& {Vargas
  Dom\'{\i}nguez}}{{Campos Rozo} \& {Vargas Dom\'{\i}nguez}}{2015}]{campos2015}
{Campos Rozo} J.~I.,  {Vargas Dom\'{\i}nguez} S.,  2015, AGU Fall Meeting
  Abstracts, \href {http://adsabs.harvard.edu/abs/2015AGUFMSH43B2443C} {}

\bibitem[\protect\citeauthoryear{{Carlin}, {Asensio Ramos}  \& {Trujillo
  Bueno}}{{Carlin} et~al.}{2013}]{carlin2013}
{Carlin} E.~S.,  {Asensio Ramos} A.,   {Trujillo Bueno} J.,  2013, \mn@doi
  [\apj] {10.1088/0004-637X/764/1/40}, \href
  {http://adsabs.harvard.edu/abs/2013ApJ...764...40C} {764, 40}

\bibitem[\protect\citeauthoryear{{Carlsson}, {Hansteen}, {Gudiksen},
  {Leenaarts}  \& {De Pontieu}}{{Carlsson} et~al.}{2016}]{Carlsson2016}
{Carlsson} M.,  {Hansteen} V.~H.,  {Gudiksen} B.~V.,  {Leenaarts} J.,   {De
  Pontieu} B.,  2016, \mn@doi [\aap] {10.1051/0004-6361/201527226}, \href
  {http://adsabs.harvard.edu/abs/2016A\%26A...585A...4C} {585, A4}

\bibitem[\protect\citeauthoryear{{Casini}, {L\'{o}pez Ariste}, {Tomczyk}  \&
  {Lites}}{{Casini} et~al.}{2003}]{Casini2003}
{Casini} R.,  {L\'{o}pez Ariste} A.,  {Tomczyk} S.,   {Lites} B.~W.,  2003,
  \mn@doi [\apjl] {10.1086/380496}, \href
  {http://adsabs.harvard.edu/abs/2003ApJ...598L..67C} {598, L67}

\bibitem[\protect\citeauthoryear{{Cheung}, {Sch\"{u}ssler}  \&
  {Moreno-Insertis}}{{Cheung} et~al.}{2007}]{cheung2007}
{Cheung} M.~C.~M.,  {Sch\"{u}ssler} M.,   {Moreno-Insertis} F.,  2007, \mn@doi
  [\aap] {10.1051/0004-6361:20066390}, \href
  {http://adsabs.harvard.edu/abs/2007A\%26A...461.1163C} {461, 1163}

\bibitem[\protect\citeauthoryear{{Cox}}{{Cox}}{2000}]{Cox2000}
{Cox} A.~N.,  2000, {Allen's astrophysical quantities}

\bibitem[\protect\citeauthoryear{{De Pontieu} et~al.,}{{De Pontieu}
  et~al.}{2014}]{IRIS2014}
{De Pontieu} B.,  et~al., 2014, \mn@doi [\solphys] {10.1007/s11207-014-0485-y},
  \href {http://adsabs.harvard.edu/abs/2014SoPh..289.2733D} {289, 2733}

\bibitem[\protect\citeauthoryear{{Diercke}, {Kuckein}, {Verma}  \&
  {Denker}}{{Diercke} et~al.}{2018}]{Diercke2018}
{Diercke} A.,  {Kuckein} C.,  {Verma} M.,   {Denker} C.,  2018, \mn@doi [\aap]
  {10.1051/0004-6361/201730536}, \href
  {http://adsabs.harvard.edu/abs/2018A\%26A...611A..64D} {611, A64}

\bibitem[\protect\citeauthoryear{{Gudiksen}, {Carlsson}, {Hansteen}, {Hayek},
  {Leenaarts}  \& {Mart\'{\i}nez-Sykora}}{{Gudiksen}
  et~al.}{2011}]{Gudiksen2011}
{Gudiksen} B.~V.,  {Carlsson} M.,  {Hansteen} V.~H.,  {Hayek} W.,  {Leenaarts}
  J.,   {Mart\'{\i}nez-Sykora} J.,  2011, \mn@doi [\aap]
  {10.1051/0004-6361/201116520}, \href
  {http://adsabs.harvard.edu/abs/2011A\%26A...531A.154G} {531, A154}

\bibitem[\protect\citeauthoryear{{Gun\'{a}r} \& {Mackay}}{{Gun\'{a}r} \&
  {Mackay}}{2016}]{Gunar2016}
{Gun\'{a}r} S.,  {Mackay} D.~H.,  2016, \mn@doi [\aap]
  {10.1051/0004-6361/201527704}, \href
  {http://adsabs.harvard.edu/abs/2016A\%26A...592A..60G} {592, A60}

\bibitem[\protect\citeauthoryear{{Heinzel}}{{Heinzel}}{2007}]{heinzel2007}
{Heinzel} P.,  2007, in {Heinzel} P.,  {Dorotovi{\v c}} I.,   {Rutten} R.~J.,
  eds,  {Astronomical Society of the Pacific Conference Series} Vol. 368, {The
  Physics of Chromospheric Plasmas}. p. Heinzel (\mn@eprint {arXiv}
  {0705.1464})

\bibitem[\protect\citeauthoryear{{Hill}, {Fischer}, {Grier}, {Leibacher},
  {Jones}, {Jones}, {Kupke}  \& {Stebbins}}{{Hill} et~al.}{1994}]{hill1994}
{Hill} F.,  {Fischer} G.,  {Grier} J.,  {Leibacher} J.~W.,  {Jones} H.~B.,
  {Jones} P.~P.,  {Kupke} R.,   {Stebbins} R.~T.,  1994, \mn@doi [\solphys]
  {10.1007/BF00680443}, \href
  {http://cdsads.u-strasbg.fr/abs/1994SoPh..152..321H} {152, 321}

\bibitem[\protect\citeauthoryear{{Keppens} \& {Xia}}{{Keppens} \&
  {Xia}}{2014}]{Keppens2014}
{Keppens} R.,  {Xia} C.,  2014, \mn@doi [\apj] {10.1088/0004-637X/789/1/22},
  \href {http://adsabs.harvard.edu/abs/2014ApJ...789...22K} {789, 22}

\bibitem[\protect\citeauthoryear{{Khomenko}, {D{\'{\i}}az}, {de Vicente},
  {Collados}  \& {Luna}}{{Khomenko} et~al.}{2014}]{Khomenko2014RTI}
{Khomenko} E.,  {D{\'{\i}}az} A.,  {de Vicente} A.,  {Collados} M.,   {Luna}
  M.,  2014, \mn@doi [\aap] {10.1051/0004-6361/201322918}, \href
  {http://adsabs.harvard.edu/abs/2014A%26A...565A..45K} {565, A45}

\bibitem[\protect\citeauthoryear{{Khomenko}, {Collados}  \&
  {D\'{\i}az}}{{Khomenko} et~al.}{2016}]{Khomenko2016}
{Khomenko} E.,  {Collados} M.,   {D\'{\i}az} A.~J.,  2016, \mn@doi [\apj]
  {10.3847/0004-637X/823/2/132}, \href
  {http://adsabs.harvard.edu/abs/2016ApJ...823..132K} {823, 132}

\bibitem[\protect\citeauthoryear{{Kippenhahn} \& {Schl\"{u}ter}}{{Kippenhahn}
  \& {Schl\"{u}ter}}{1957}]{Kippenhahn1957}
{Kippenhahn} R.,  {Schl\"{u}ter} A.,  1957, \zap, \href
  {http://adsabs.harvard.edu/abs/1957ZA.....43...36K} {43, 36}

\bibitem[\protect\citeauthoryear{{Kuckein}, {Mart\'{\i}nez Pillet}  \&
  {Centeno}}{{Kuckein} et~al.}{2012}]{kuckein2012}
{Kuckein} C.,  {Mart\'{\i}nez Pillet} V.,   {Centeno} R.,  2012, \mn@doi [\aap]
  {10.1051/0004-6361/201117675}, \href
  {http://adsabs.harvard.edu/abs/2012A\%26A...539A.131K} {539, A131}

\bibitem[\protect\citeauthoryear{{Kuckein}, {Verma}  \& {Denker}}{{Kuckein}
  et~al.}{2016}]{kuckein2016}
{Kuckein} C.,  {Verma} M.,   {Denker} C.,  2016, \mn@doi [\aap]
  {10.1051/0004-6361/201526636}, \href
  {http://adsabs.harvard.edu/abs/2016A\%26A...589A..84K} {589, A84}

\bibitem[\protect\citeauthoryear{{Leenaarts} \& {Carlsson}}{{Leenaarts} \&
  {Carlsson}}{2009}]{Leenaarts2009M}
{Leenaarts} J.,  {Carlsson} M.,  2009, in {Lites} B.,  {Cheung} M.,  {Magara}
  T.,  {Mariska} J.,   {Reeves} K.,  eds,  Astronomical Society of the Pacific
  Conference Series Vol. 415, The Second Hinode Science Meeting: Beyond
  Discovery-Toward Understanding. p.~87

\bibitem[\protect\citeauthoryear{{Leenaarts}, {de la Cruz Rodr\'{\i}guez},
  {Kochukhov}  \& {Carlsson}}{{Leenaarts} et~al.}{2014}]{leenaarts2014}
{Leenaarts} J.,  {de la Cruz Rodr\'{\i}guez} J.,  {Kochukhov} O.,   {Carlsson}
  M.,  2014, \mn@doi [\apjl] {10.1088/2041-8205/784/1/L17}, \href
  {http://adsabs.harvard.edu/abs/2014ApJ...784L..17L} {784, L17}

\bibitem[\protect\citeauthoryear{{Lemen} et~al.,}{{Lemen}
  et~al.}{2012}]{aia2012}
{Lemen} J.~R.,  et~al., 2012, \mn@doi [\solphys] {10.1007/s11207-011-9776-8},
  \href {http://adsabs.harvard.edu/abs/2012SoPh..275...17L} {275, 17}

\bibitem[\protect\citeauthoryear{{Leroy}, {Ratier}  \& {Bommier}}{{Leroy}
  et~al.}{1977}]{Leroy1977}
{Leroy} J.~L.,  {Ratier} G.,   {Bommier} V.,  1977, \aap, \href
  {http://adsabs.harvard.edu/abs/1977A\%26A....54..811L} {54, 811}

\bibitem[\protect\citeauthoryear{{Lin}, {Engvold}  \& {Wiik}}{{Lin}
  et~al.}{2003}]{Lin2003}
{Lin} Y.,  {Engvold} O.~R.,   {Wiik} J.~E.,  2003, \mn@doi [\solphys]
  {10.1023/A:1026150809598}, \href
  {http://adsabs.harvard.edu/abs/2003SoPh..216..109L} {216, 109}

\bibitem[\protect\citeauthoryear{{Lites}}{{Lites}}{2005}]{Lites2005}
{Lites} B.~W.,  2005, \mn@doi [\apj] {10.1086/428080}, \href
  {http://adsabs.harvard.edu/abs/2005ApJ...622.1275L} {622, 1275}

\bibitem[\protect\citeauthoryear{{L{\'o}pez Ariste} \& {Casini}}{{L{\'o}pez
  Ariste} \& {Casini}}{2002}]{LopezAriste2002}
{L{\'o}pez Ariste} A.,  {Casini} R.,  2002, \mn@doi [\apj] {10.1086/341260},
  \href {http://adsabs.harvard.edu/abs/2002ApJ...575..529L} {575, 529}

\bibitem[\protect\citeauthoryear{{Luna}, {Moreno-Insertis}  \& {Priest}}{{Luna}
  et~al.}{2015}]{Luna2015}
{Luna} M.,  {Moreno-Insertis} F.,   {Priest} E.,  2015, \mn@doi [\apjl]
  {10.1088/2041-8205/808/1/L23}, \href
  {http://adsabs.harvard.edu/abs/2015ApJ...808L..23L} {808, L23}

\bibitem[\protect\citeauthoryear{{Mackay}, {Karpen}, {Ballester}, {Schmieder}
  \& {Aulanier}}{{Mackay} et~al.}{2010}]{mackay2010}
{Mackay} D.~H.,  {Karpen} J.~T.,  {Ballester} J.~L.,  {Schmieder} B.,
  {Aulanier} G.,  2010, \mn@doi [\ssr] {10.1007/s11214-010-9628-0}, \href
  {http://adsabs.harvard.edu/abs/2010SSRv..151..333M} {151, 333}

\bibitem[\protect\citeauthoryear{{Manso Sainz} \& {Trujillo Bueno}}{{Manso
  Sainz} \& {Trujillo Bueno}}{2010}]{manso2010}
{Manso Sainz} R.,  {Trujillo Bueno} J.,  2010, \mn@doi [\apj]
  {10.1088/0004-637X/722/2/1416}, \href
  {http://adsabs.harvard.edu/abs/2010ApJ...722.1416M} {722, 1416}

\bibitem[\protect\citeauthoryear{{Martens} \& {Zwaan}}{{Martens} \&
  {Zwaan}}{2001}]{Martens2001}
{Martens} P.~C.,  {Zwaan} C.,  2001, \mn@doi [\apj] {10.1086/322279}, \href
  {http://adsabs.harvard.edu/abs/2001ApJ...558..872M} {558, 872}

\bibitem[\protect\citeauthoryear{{Mart\'{\i}nez Gonz\'{a}lez}, {Manso Sainz},
  {Asensio Ramos}  \& {Belluzzi}}{{Mart\'{\i}nez Gonz\'{a}lez}
  et~al.}{2012}]{marian2012}
{Mart\'{\i}nez Gonz\'{a}lez} M.~J.,  {Manso Sainz} R.,  {Asensio Ramos} A.,
  {Belluzzi} L.,  2012, \mn@doi [\mnras] {10.1111/j.1365-2966.2011.19681.x},
  \href {http://adsabs.harvard.edu/abs/2012MNRAS.419..153M} {419, 153}

\bibitem[\protect\citeauthoryear{{Mart\'{\i}nez Gonz\'{a}lez}, {Manso Sainz},
  {Asensio Ramos}, {Beck}, {de la Cruz Rodr\'{\i}guez}  \&
  {D\'{\i}az}}{{Mart\'{\i}nez Gonz\'{a}lez} et~al.}{2015}]{marian2015}
{Mart\'{\i}nez Gonz\'{a}lez} M.~J.,  {Manso Sainz} R.,  {Asensio Ramos} A.,
  {Beck} C.,  {de la Cruz Rodr\'{\i}guez} J.,   {D\'{\i}az} A.~J.,  2015,
  \mn@doi [\apj] {10.1088/0004-637X/802/1/3}, \href
  {http://adsabs.harvard.edu/abs/2015ApJ...802....3M} {802, 3}

\bibitem[\protect\citeauthoryear{{Neckel} \& {Labs}}{{Neckel} \&
  {Labs}}{1984}]{neckel1984}
{Neckel} H.,  {Labs} D.,  1984, \mn@doi [\solphys] {10.1007/BF00173953}, \href
  {http://adsabs.harvard.edu/abs/1984SoPh...90..205N} {90, 205}

\bibitem[\protect\citeauthoryear{{Okamoto}, {Liu}  \& {Tsuneta}}{{Okamoto}
  et~al.}{2016}]{Okamoto2016}
{Okamoto} T.~J.,  {Liu} W.,   {Tsuneta} S.,  2016, \mn@doi [\apj]
  {10.3847/0004-637X/831/2/126}, \href
  {http://adsabs.harvard.edu/abs/2016ApJ...831..126O} {831, 126}

\bibitem[\protect\citeauthoryear{{Orozco Su\'{a}rez}, {Asensio Ramos}  \&
  {Trujillo Bueno}}{{Orozco Su\'{a}rez} et~al.}{2014}]{orozco2014}
{Orozco Su\'{a}rez} D.,  {Asensio Ramos} A.,   {Trujillo Bueno} J.,  2014,
  \mn@doi [\aap] {10.1051/0004-6361/201322903}, \href
  {http://adsabs.harvard.edu/abs/2014A\%26A...566A..46O} {566, A46}

\bibitem[\protect\citeauthoryear{{Panesar}}{{Panesar}}{2014}]{Panesar2014}
{Panesar} N.~K.,  2014, PhD thesis, Georg-August-Universit{\"a}t G{\"o}ttingen,
  Institut f{\"u}r Astrophysik, Germany, \mn@doi{10.5281/zenodo.581205}

\bibitem[\protect\citeauthoryear{{Park} et~al.,}{{Park}
  et~al.}{2013}]{Park2013}
{Park} H.,  et~al., 2013, \mn@doi [\solphys] {10.1007/s11207-013-0271-2}, \href
  {http://adsabs.harvard.edu/abs/2013SoPh..288..105P} {288, 105}

\bibitem[\protect\citeauthoryear{{Pesnell}, {Thompson}  \&
  {Chamberlin}}{{Pesnell} et~al.}{2012}]{sdo2012}
{Pesnell} W.~D.,  {Thompson} B.~J.,   {Chamberlin} P.~C.,  2012, \mn@doi
  [\solphys] {10.1007/s11207-011-9841-3}, \href
  {http://adsabs.harvard.edu/abs/2012SoPh..275....3P} {275, 3}

\bibitem[\protect\citeauthoryear{{Quintero Noda}, {Shimizu}, {de la Cruz
  Rodr\'{\i}guez}, {Katsukawa}, {Ichimoto}, {Anan}  \& {Suematsu}}{{Quintero
  Noda} et~al.}{2016}]{Quintero2016}
{Quintero Noda} C.,  {Shimizu} T.,  {de la Cruz Rodr\'{\i}guez} J.,
  {Katsukawa} Y.,  {Ichimoto} K.,  {Anan} T.,   {Suematsu} Y.,  2016, \mn@doi
  [\mnras] {10.1093/mnras/stw867}, \href
  {http://adsabs.harvard.edu/abs/2016MNRAS.459.3363Q} {459, 3363}

\bibitem[\protect\citeauthoryear{{Ruiz Cobo} \& {del Toro Iniesta}}{{Ruiz Cobo}
  \& {del Toro Iniesta}}{1992}]{RuizCobo1992}
{Ruiz Cobo} B.,  {del Toro Iniesta} J.~C.,  1992, \mn@doi [\apj]
  {10.1086/171862}, \href {http://adsabs.harvard.edu/abs/1992ApJ...398..375R}
  {398, 375}

\bibitem[\protect\citeauthoryear{{Schad}, {Penn}, {Lin}  \& {Judge}}{{Schad}
  et~al.}{2016}]{schad2016}
{Schad} T.~A.,  {Penn} M.~J.,  {Lin} H.,   {Judge} P.~G.,  2016, \mn@doi [\apj]
  {10.3847/0004-637X/833/1/5}, \href
  {http://adsabs.harvard.edu/abs/2016ApJ...833....5S} {833, 5}

\bibitem[\protect\citeauthoryear{{Scharmer}, {Bjelksjo}, {Korhonen}, {Lindberg}
   \& {Petterson}}{{Scharmer} et~al.}{2003}]{scharmer2003}
{Scharmer} G.~B.,  {Bjelksjo} K.,  {Korhonen} T.~K.,  {Lindberg} B.,
  {Petterson} B.,  2003, in {Keil} S.~L.,  {Avakyan} S.~V.,  eds, ~{\procspie}
  Vol. 4853, {Innovative Telescopes and Instrumentation for Solar
  Astrophysics}. pp 341--350, \mn@doi{10.1117/12.460377}

\bibitem[\protect\citeauthoryear{{Scharmer} et~al.,}{{Scharmer}
  et~al.}{2008}]{scharmer2008}
{Scharmer} G.~B.,  et~al., 2008, \mn@doi [\apjl] {10.1086/595744}, \href
  {http://adsabs.harvard.edu/abs/2008ApJ...689L..69S} {689, L69}

\bibitem[\protect\citeauthoryear{{Schmieder}, {Tziotziou}  \&
  {Heinzel}}{{Schmieder} et~al.}{2003}]{Schmieder2003}
{Schmieder} B.,  {Tziotziou} K.,   {Heinzel} P.,  2003, \mn@doi [\aap]
  {10.1051/0004-6361:20030126}, \href
  {http://adsabs.harvard.edu/abs/2003A\%26A...401..361S} {401, 361}

\bibitem[\protect\citeauthoryear{{Socas-Navarro}}{{Socas-Navarro}}{2005}]{2005ApJ...633L..57S}
{Socas-Navarro} H.,  2005, \mn@doi [\apjl] {10.1086/498145}, \href
  {http://adsabs.harvard.edu/abs/2005ApJ...633L..57S} {633, L57}

\bibitem[\protect\citeauthoryear{{Socas-Navarro}, {Trujillo Bueno}  \& {Ruiz
  Cobo}}{{Socas-Navarro} et~al.}{2000}]{socas2000}
{Socas-Navarro} H.,  {Trujillo Bueno} J.,   {Ruiz Cobo} B.,  2000, \mn@doi
  [\apj] {10.1086/308414}, \href
  {http://adsabs.harvard.edu/abs/2000ApJ...530..977S} {530, 977}

\bibitem[\protect\citeauthoryear{{Socas-Navarro}, {de la Cruz Rodr\'{\i}guez},
  {Asensio Ramos}, {Trujillo Bueno}  \& {Ruiz Cobo}}{{Socas-Navarro}
  et~al.}{2015}]{SocasNavarro2015}
{Socas-Navarro} H.,  {de la Cruz Rodr\'{\i}guez} J.,  {Asensio Ramos} A.,
  {Trujillo Bueno} J.,   {Ruiz Cobo} B.,  2015, \mn@doi [\aap]
  {10.1051/0004-6361/201424860}, \href
  {http://adsabs.harvard.edu/abs/2015A\%26A...577A...7S} {577, A7}

\bibitem[\protect\citeauthoryear{{Stellmacher}, {Wiehr}  \&
  {Dammasch}}{{Stellmacher} et~al.}{2003}]{Stellmacher2003}
{Stellmacher} G.,  {Wiehr} E.,   {Dammasch} I.~E.,  2003, \mn@doi [\solphys]
  {10.1023/A:1027310303994}, \href
  {http://adsabs.harvard.edu/abs/2003SoPh..217..133S} {217, 133}

\bibitem[\protect\citeauthoryear{{SunPy Community} et~al.,}{{SunPy Community}
  et~al.}{2015}]{sunpy2015}
{SunPy Community} T.,  et~al., 2015, \mn@doi [Computational Science and
  Discovery] {10.1088/1749-4699/8/1/014009}, \href
  {http://adsabs.harvard.edu/abs/2015CS\%26D....8a4009S} {8, 014009}

\bibitem[\protect\citeauthoryear{{Tziotziou}}{{Tziotziou}}{2007}]{Tziotziou2007}
{Tziotziou} K.,  2007, in {Heinzel} P.,  {Dorotovi{\v c}} I.,   {Rutten} R.~J.,
   eds,  {Astronomical Society of the Pacific Conference Series} Vol. 368, {The
  Physics of Chromospheric Plasmas}. p. Heinzel (\mn@eprint {arXiv}
  {0704.1558})

\bibitem[\protect\citeauthoryear{{Uitenbroek}}{{Uitenbroek}}{1989}]{Uitenbroek1989}
{Uitenbroek} H.,  1989, \aap, \href
  {http://adsabs.harvard.edu/abs/1989A\%26A...213..360U} {213, 360}

\bibitem[\protect\citeauthoryear{{Uitenbroek}}{{Uitenbroek}}{2001}]{Uitenbroek2001}
{Uitenbroek} H.,  2001, \mn@doi [\apj] {10.1086/321659}, \href
  {http://adsabs.harvard.edu/abs/2001ApJ...557..389U} {557, 389}

\bibitem[\protect\citeauthoryear{{Vecchio}, {Cauzzi}, {Reardon}, {Janssen}  \&
  {Rimmele}}{{Vecchio} et~al.}{2007}]{Vecchio2007}
{Vecchio} A.,  {Cauzzi} G.,  {Reardon} K.~P.,  {Janssen} K.,   {Rimmele} T.,
  2007, \mn@doi [\aap] {10.1051/0004-6361:20066415}, \href
  {http://adsabs.harvard.edu/abs/2007A\%26A...461L...1V} {461, L1}

\bibitem[\protect\citeauthoryear{{Wedemeyer-B\"{o}hm} \&
  {Carlsson}}{{Wedemeyer-B\"{o}hm} \& {Carlsson}}{2011}]{Wedemeyer2011}
{Wedemeyer-B\"{o}hm} S.,  {Carlsson} M.,  2011, \mn@doi [\aap]
  {10.1051/0004-6361/201016186}, \href
  {http://adsabs.harvard.edu/abs/2011A\%26A...528A...1W} {528, A1}

\bibitem[\protect\citeauthoryear{{Yi} \& {Molowny-Horas}}{{Yi} \&
  {Molowny-Horas}}{1995}]{yi1995}
{Yi} Z.,  {Molowny-Horas} R.,  1995, \aap, \href
  {http://adsabs.harvard.edu/abs/1995A\%26A...295..199Y} {295, 199}

\bibitem[\protect\citeauthoryear{Zirker, Engvold  \& Martin}{Zirker
  et~al.}{1998}]{Zirker1998}
Zirker J.~B.,  Engvold O.,   Martin S.~F.,  1998, \mn@doi [Nature]
  {10.1038/24798}, 396, 440

\bibitem[\protect\citeauthoryear{{de la Cruz Rodr\'{\i}guez} \& {Piskunov}}{{de
  la Cruz Rodr\'{\i}guez} \& {Piskunov}}{2013}]{delaCruz2013d}
{de la Cruz Rodr\'{\i}guez} J.,  {Piskunov} N.,  2013, \mn@doi [\apj]
  {10.1088/0004-637X/764/1/33}, \href
  {http://adsabs.harvard.edu/abs/2013ApJ...764...33D} {764, 33}

\bibitem[\protect\citeauthoryear{{de la Cruz Rodr\'{\i}guez}, {Socas-Navarro},
  {Carlsson}  \& {Leenaarts}}{{de la Cruz Rodr\'{\i}guez}
  et~al.}{2012}]{delaCruz2012}
{de la Cruz Rodr\'{\i}guez} J.,  {Socas-Navarro} H.,  {Carlsson} M.,
  {Leenaarts} J.,  2012, \mn@doi [\aap] {10.1051/0004-6361/201218825}, \href
  {http://adsabs.harvard.edu/abs/2012A\%26A...543A..34D} {543, A34}

\bibitem[\protect\citeauthoryear{{de la Cruz Rodr\'{\i}guez}, {Rouppe van der
  Voort}, {Socas-Navarro}  \& {van Noort}}{{de la Cruz Rodr\'{\i}guez}
  et~al.}{2013a}]{delaCruz2013}
{de la Cruz Rodr\'{\i}guez} J.,  {Rouppe van der Voort} L.,  {Socas-Navarro}
  H.,   {van Noort} M.,  2013a, \mn@doi [\aap] {10.1051/0004-6361/201321629},
  \href {http://adsabs.harvard.edu/abs/2013A\%26A...556A.115D} {556, A115}

\bibitem[\protect\citeauthoryear{{de la Cruz Rodr\'{\i}guez}, {De Pontieu},
  {Carlsson}  \& {Rouppe van der Voort}}{{de la Cruz Rodr\'{\i}guez}
  et~al.}{2013b}]{delaCruz2013h}
{de la Cruz Rodr\'{\i}guez} J.,  {De Pontieu} B.,  {Carlsson} M.,   {Rouppe van
  der Voort} L.~H.~M.,  2013b, \mn@doi [\apjl] {10.1088/2041-8205/764/1/L11},
  \href {http://adsabs.harvard.edu/abs/2013ApJ...764L..11D} {764, L11}

\bibitem[\protect\citeauthoryear{{de la Cruz Rodr\'{\i}guez}, {L\"{o}fdahl},
  {S\"{u}tterlin}, {Hillberg}  \& {Rouppe van der Voort}}{{de la Cruz
  Rodr\'{\i}guez} et~al.}{2015a}]{delaCruz2015}
{de la Cruz Rodr\'{\i}guez} J.,  {L\"{o}fdahl} M.~G.,  {S\"{u}tterlin} P.,
  {Hillberg} T.,   {Rouppe van der Voort} L.,  2015a, \mn@doi [\aap]
  {10.1051/0004-6361/201424319}, \href
  {http://adsabs.harvard.edu/abs/2015A\%26A...573A..40D} {573, A40}

\bibitem[\protect\citeauthoryear{{de la Cruz Rodr{\'{\i}}guez}, {Hansteen},
  {Bellot-Rubio}  \& {Ortiz}}{{de la Cruz Rodr{\'{\i}}guez}
  et~al.}{2015b}]{2015ApJ...810..145D}
{de la Cruz Rodr{\'{\i}}guez} J.,  {Hansteen} V.,  {Bellot-Rubio} L.,   {Ortiz}
  A.,  2015b, \mn@doi [\apj] {10.1088/0004-637X/810/2/145}, \href
  {http://adsabs.harvard.edu/abs/2015ApJ...810..145D} {810, 145}

\bibitem[\protect\citeauthoryear{{{\v S}t{\v e}p{\'a}n} \& {Trujillo
  Bueno}}{{{\v S}t{\v e}p{\'a}n} \& {Trujillo Bueno}}{2016}]{Stephan2016}
{{\v S}t{\v e}p{\'a}n} J.,  {Trujillo Bueno} J.,  2016, \mn@doi [\apjl]
  {10.3847/2041-8205/826/1/L10}, \href
  {http://adsabs.harvard.edu/abs/2016ApJ...826L..10S} {826, L10}

\bibitem[\protect\citeauthoryear{{van Ballegooijen} \& {Cranmer}}{{van
  Ballegooijen} \& {Cranmer}}{2010}]{vanB2010}
{van Ballegooijen} A.~A.,  {Cranmer} S.~R.,  2010, \mn@doi [\apj]
  {10.1088/0004-637X/711/1/164}, \href
  {http://adsabs.harvard.edu/abs/2010ApJ...711..164V} {711, 164}

\bibitem[\protect\citeauthoryear{{van Ballegooijen} \& {Martens}}{{van
  Ballegooijen} \& {Martens}}{1989}]{van1989}
{van Ballegooijen} A.~A.,  {Martens} P.~C.~H.,  1989, \mn@doi [\apj]
  {10.1086/167766}, \href {http://adsabs.harvard.edu/abs/1989ApJ...343..971V}
  {343, 971}

\bibitem[\protect\citeauthoryear{{van Noort}, {Rouppe van der Voort}  \&
  {L\"{o}fdahl}}{{van Noort} et~al.}{2005}]{vanNoort2005}
{van Noort} M.,  {Rouppe van der Voort} L.,   {L\"{o}fdahl} M.~G.,  2005,
  \mn@doi [\solphys] {10.1007/s11207-005-5782-z}, \href
  {http://adsabs.harvard.edu/abs/2005SoPh..228..191V} {228, 191}

\makeatother
\end{thebibliography}

\end{document}